\documentclass[aps,pra,showpacs,twocolumn]{revtex4}

\usepackage{amsmath}
\usepackage{amssymb}
\usepackage{graphicx}
\usepackage{subfig}
\usepackage{color}

\begin{document}

\title{CLASSICAL DYNAMICS OF A THIN MOVING MIRROR INTERACTING WITH A LASER}

\author{L. O. Casta\~nos$^{*}$ and R. Weder$^{+}$}

\affiliation{Departamento de F\'{i}sica Matem\'{a}tica, Instituto de Investigaciones en Matem\'{a}ticas Aplicadas y en Sistemas, Universidad Nacional Aut\'{o}noma de M\'{e}xico, Apartado Postal 20-726, M\'{e}xico DF 01000, M\'{e}xico}

\begin{abstract}
We analyze the classical dynamics of a system composed of a one-dimensional cavity with a perfect, fixed mirror and a movable mirror with non-zero transparency interacting with a monochromatic laser. The movable mirror can deviate far from an equilibrium position, it is assumed to be thin so that it is modelled by a delta function, and we use the exact modes of the complete system. The transparency and the mirror-position dependent cavity resonance frequencies are built into the modes and this allows us to deduce that the radiation pressure force comes from a periodic potential with period half the wavelength of the field. The exact modes and the radiation pressure potential allow us to give intuitive physical interpretations of the dynamics of the system and to obtain approximate analytic solutions for the motion of the mirror. Three regimes are identified depending on the intensity of the field and it is found that the dynamics can be qualitatively very different in each of them with some regimes being very sensitive to the values of the parameters. Moreover, we determine conditions when the Maxwell-Newton equations used to describe the system constitute accurate approximations of the exact equations governing the dynamics.    
\end{abstract}

\maketitle


\section{INTRODUCTION}

The area of optomechanics has received much attention due to potential applications ranging from precision measurements to fundamental tests of quantum mechanics to optical communications \cite{Trends,Cold, Israel0}. Optomechanical systems not only have the potential to become a powerful probe with which to explore the quantum world \cite{Cold,Mar0,Restrepo,HighFreq,MM,NoClasico,PhotonShuttle,TwoMode}, but are also of great interest in the area of optical micro-electromechanical systems \cite{Israel0} and constitute a setting in which classical non-linear dynamics have to be analyzed \cite{Kip1,Kip2,Mar1,Mar2,Italia1,Italia2,Israel1,Israel2,Nonlinear}.

One of the paradigmatic models in optomechanics consists of a cavity with one movable mirror. The movable mirror is a mechanical oscillator and it is coupled to the electromagnetic field by radiation pressure and by thermal or bolometric effects (absorption of photons distorting and displacing the mirror). The classical dynamics of this type of systems has been studied in \cite{Kip1,Kip2,Mar1,Mar2,Italia1,Italia2,Israel1,Israel2,Nonlinear}. For the quantum dynamics we refer the reader to \cite{Trends,Cold,Mar0,Restrepo,HighFreq,MM,NoClasico,PhotonShuttle,TwoMode}.

The classical dynamics of an optomechanical set-up similar to the model described above and dominated by bolometric forces has been studied both experimentally and theoretically \cite{Mar1,Mar2}. It was found that self-sustained oscillations of the movable mirror are present, that their amplitude settles into one of several attractors (see also \cite{Nonlinear} for this phenomenon), and that there is a regime where two mechanical modes of oscillation can be excited. Moreover, it was suggested that the multi-stability can be used as a device to measure small displacements. 

The classical dynamics of a set-up dominated by radiation pressure has also been studied both experimentally and theoretically \cite{Kip1,Kip2}. It is composed of a toroid cavity and a pump beam. The toroid cavity executes a periodic motion that was explained by the following intuitive physical picture: radiation pressure induces a flex of the cavity that takes it out of resonance with the pump beam; this leads to a reduction of radiation pressure and, as a consequence, the cavity moves back into resonance and the process starts over again. It was found that the cavity vibrations cause modulation of the input pump beam and that this modulation becomes random oscillations with sufficiently high pump power \cite{Kip1}. Moreover, a radiation pressure induced parametric oscillation instability (a regenerative oscillation of mechanical eigenmodes) was analyzed \cite{Kip2}.  

The case where both radiation pressure and thermal effects are relevant has also been studied  \cite{Italia1,Italia2,Israel1,Israel2}. It has been shown experimentally and theoretically \cite{Italia1,Italia2} that these two types of forces operate at different time scales and can lead to dynamics in which chaotic attractors appear. Also, \cite{Israel1} established a very detailed phenomenological model for a micromechanical mirror that incorporates non-linear elastic and dissipative terms, an external excitation force as well as the radiation pressure and thermal forces. The model was used to describe the results of an experimental study of the dynamics of an optomechanical cavity with micro-mechanical mirrors in two different geometries \cite{Israel2}.

In this article we investigate the classical dynamics of a thin cavity movable mirror modelled by a delta-function interacting with a laser by means of radiation pressure.
The purpose of the article is to investigate the dynamics of the system by considering only the leading term in the force affecting the motion of the movable mirror. This is similar to the way light forces on atoms are discussed in some treatments like \cite{Cohen} where the atom is first assumed to be instantaneously fixed and the dissipative (or radiation pressure) and the reactive (or dipole) forces are obtained.
Our approach differs from most articles in that we start from approximate Maxwell-Newton equations valid when the velocity and acceleration of the movable mirror are small so that the movable mirror can deviate far from an equilibrium position. Moreover, we consider the exact modes of the complete system, that is, we do not divide them into modes of the cavity and modes outside of it as if the cavity where composed of perfect mirrors and then couple them through a phenomenological interaction \cite{Dutra}. This allows us to incorporate both the transparency of the mirror and the mirror-position dependent cavity resonance frequencies directly in the modes.

As a result of our treatment, we derive the radiation pressure force from a periodic potential and the physical process governing the dynamics in the case of the toroid cavity mentioned above is explicitly and rigorously derived from our equations. The radiation pressure potential gives physical insight into the dynamics of the system and allows us to determine approximate analytical solutions for the motion of the movable mirror. Also, special attention is paid in establishing the conditions under which the model is valid. The intention is to shed more light into the physics and the intricate dynamics of the movable mirror-electromagnetic field system as well as to have a more profound understanding of when the approximate Maxwell-Newton equations normally used to describe this type of systems are valid. 

The model of a \textit{fixed very thin mirror} with non-zero transparency modelled by a delta-function (known as the Lang-Scully-Lamb or LSL model) has been used in the past \cite{L,LSL,Nussenzveig,Penaforte,Guedes,Guedes2,Barnett,Suttorp}. Originally it was introduced as a simple more realistic model of a laser with the purpose to explain physical phenomena that could not be answered (at least satisfactorily) with the phenomenological models. In particular, it was used to explain why the laser line is so narrow, to form a better conceptual picture of the nature of a laser mode, and to account for radiation losses through one of the mirrors of the laser without the use of phenomenological models \cite{Dutra,L,LSL,Nussenzveig}. The investigations were initially restricted to semi-classical treatments and were later extended to full quantum treatments \cite{Penaforte,Guedes,Guedes2}. Afterwards, the same model was used to explain why the phenomenological models work in the good cavity limit \cite{Barnett} and to deduce a master equation to incorporate the effects of a finite mirror transmissivity \cite{Suttorp}. 

After this article was completed we learned about references \cite{DomokosI,DomokosII}. Reference \cite{DomokosI}  presents a scattering approach to describe the coupling of the electromagnetic field to a mobile scatterer (a mirror or atom) moving with constant velocity and with a dispersive dielectric constant. In particular, they derive the friction force and the diffusion affecting the scatterer with the rotating-wave-approximation. Also, \cite{DomokosII} uses a similar scattering approach to study the dynamics of a one-dimensional optical lattice including non-linearities caused by multiple reflections of photons.  

The article is organized as follows. In Section II we establish the system under study and the model used to describe it. In Section III we restrict to a single-mode field and derive the potential associated with the radiation pressure force. In Section IV we investigate the dynamics of the movable mirror. In Sections V and VI we consider the cases where the movable mirror is also subject to friction and to a harmonic oscillator potential. Finally, the conclusions are given in Section VII.


\section{THE MODEL}

Consider a one-dimensional cavity composed of a fixed and perfect (zero transparency) mirror and a movable mirror, both parallel to the $yz$-plane. The fixed mirror is located at $x=0$, while the movable mirror is located at \ $x = q(t) >0$ \ at time $t$ and has thickness $\delta_{\mbox{\tiny thick}}$ when it is at rest. We assume that the movable mirror is a linear, isotropic, non-magnetizable, and non-conducting (it does not contain any free charges or currents) dielectric when it is at rest. For the electromagnetic field we use the Gaussian system of units. We also assume that the electric field is polarized along the $z$-axis. This allows us to take the scalar potential equal to zero and to derive both the electric and magnetic fields from a vector potential. Explicitly, the vector potential $\mathbf{A}(x,t)$ is of the form
\begin{eqnarray}
\label{PotencialVectorial}
\mathbf{A}(x,t) &=& A_{0}(x,t)\mathbf{z} \ ,
\end{eqnarray}
while the electric $\mathbf{E}(x,t)$ and magnetic $\mathbf{B}(x,t)$ fields are given by
\begin{eqnarray}
\label{Campos}
\mathbf{E}(x,t) &=& -\frac{1}{c}\frac{\partial A_{0}}{\partial t}(x,t)\mathbf{z} \ , \cr
&& \cr
\mathbf{B}(x,t) &=& -\frac{\partial A_{0}}{\partial x}(x,t)\mathbf{y} \ .
\end{eqnarray}
Notice that we are working in the Coulomb gauge.

The equation for the vector potential is given by 
\begin{eqnarray}
\label{1}
\frac{\partial^{2}A_{0}}{\partial x^{2}}(x,t) \ = \ \frac{\epsilon\left[ x-q(t) \right]}{c^{2}}\frac{\partial^{2}A_{0}}{\partial t^{2}}(x,t) \ , 
\end{eqnarray}
for all \ $x >0$ \ and \ $t\in\mathbb{R}$. Here $\epsilon$ is the dielectric function associated with the movable mirror. Note that (\ref{1}) corresponds to the usual wave-equation for the potential with the mirror instantaneously fixed at $q(t)$.

In order to take into account the perfect and fixed mirror at \ $x=0$ \ one needs to impose the following boundary condition:
\begin{equation}
\label{4}
\frac{\partial A_{0}}{\partial t}(0+,t) \ = \ 0 \qquad (t\in\mathbb{R}).
\end{equation}
This condition comes from imposing \ $\mathbf{E}(x,t) = \mathbf{0}$ \ for \ $x<0$ \ and using the usual boundary conditions for the electromagnetic field. 

In reference \cite{Nuestro} it is shown that (\ref{1}) is obtained by posing Maxwell's equations in inertial reference frames in which the movable mirror is instantaneously at rest and then changing back to the laboratory reference frame (defined by the condition that the perfect mirror is fixed at $x=0$) using (instantaneous) Lorentz transformations and neglecting terms of order $\dot{q}(t)/c$, $\ddot{q}(t)/(c\omega_{0})$, and higher powers of them. Here $c$ is the speed of light in vacuum and $\omega_{0}$ is the characteristic frequency of the electromagnetic field. Hence, (\ref{1}) will be an accurate approximation to the exact equation governing the dynamics of $A_{0}(x,t)$ if the following two conditions are satisfied:
\begin{eqnarray}
\label{1condicionesEq}
\left\vert \frac{\dot{q}(t)}{c} \right\vert , \ \left\vert \frac{\ddot{q}(t)}{c\omega_{0}} \right\vert &\ll& 1 \qquad (t\in\mathbb{R}) \ .
\end{eqnarray}

In reference \cite{Nuestro} it is shown that equations (\ref{1}) and (\ref{4}) are valid for general dielectric functions $\epsilon[x-q(t)]$. In the rest of the article we assume that the movable mirror is very thin, that is
\begin{eqnarray}
\label{1angosto}
\delta_{\mbox{\tiny thick}} &\ll& \lambda \ \equiv \ \frac{2\pi c}{\omega_{0}}  \ .
\end{eqnarray}
Hence, one can approximate the dielectric function associated with the movable mirror by a delta function:
\begin{eqnarray}
\label{1delta}
\epsilon \left[ x-q(t) \right] &=& 1 +4\pi \chi_{0} \delta\left[ x-q(t) \right] \ .
\end{eqnarray}
Here $\chi_{0}$ has units of length. In \cite{Nuestro} we prove that (\ref{1}) is valid with $\epsilon[x-q(t)]$ given in (\ref{1delta}) by a limiting process starting with regular $\epsilon[x-q(t)]$. Notice that (\ref{1}) and (\ref{4}) with (\ref{1delta}) correspond to the LSL model discussed in the Introduction but now with a moving mirror. Moreover, we note that movable mirrors satisfying (\ref{1angosto}) have already been used experimentally in other optomechanical set-ups \cite{Harris}.

Using the law of conservation of linear momentum for the system (electromagnetic field + dielectric mirror at $x= q(t)$ + fixed perfect mirror at $x=0$) it follows from (\ref{1}) and (\ref{1delta}) that the equation for the movable mirror is given by \cite{Nuestro} 
\begin{eqnarray}
\label{2}
M_{0}\ddot{q}(t) &=& -\frac{1}{8\pi}\left\{ \ \left[ \frac{\partial A_{0}}{\partial x}[q(t)+,t] \right]^{2} \right. \cr
&& \qquad\qquad \left. - \left[  \frac{\partial A_{0}}{\partial x}[q(t)-,t] \right]^{2} \ \right\} \ ,
\end{eqnarray}
for all $t\in\mathbb{R}$. Here $M_{0}$ has units of mass per unit area and 
\begin{eqnarray}
\label{LimiteLateral}
\frac{\partial A_{0}}{\partial x}\left[ q(t)\pm , t \right] &=& \mbox{lim}_{x \rightarrow q(t)^{\pm}} \frac{\partial A_{0}}{\partial x}\left( x , t \right) \ .
\end{eqnarray}
Note that the right-hand side of (\ref{2}) is the radiation pressure exerted by the electromagnetic field on the mirror instantaneously fixed at $q(t)$.

It is important to identify which processes are not taken into account because we have only considered the leading order of the electromagnetic force acting on the movable mirror. In other words, in (\ref{1}) and (\ref{2}) we have neglected terms of order $\dot{q}(t)/c$, $\ddot{q}(t)/(c\omega_{0})$, and higher powers of them. The first corrections to the force on the right-hand side of (\ref{2}) are a friction force proportional to $\dot{q}(t)/c$ \cite{Nuestro,DomokosI} and an acceleration-dependent force proportional to $\ddot{q}(t)/(c\omega_{0})$ \cite{Nuestro}. The friction force leads to dissipation in the motion of the mirror, while the acceleration-dependent force leads a renormalized mass of the movable mirror. In addition to the friction and the acceleration-dependent forces, field amplitudes corresponding to different wave-numbers are mixed \cite{Nuestro,DomokosI}. This last process occurs when terms proportional to $\dot{q}(t)/c$ are not neglected. Equations (\ref{1}) and (\ref{2}) are accurate approximations when (\ref{1condicionesEq}) is valid, since they imply that the aforementioned friction force, modification of the mass, and mixing of wave-numbers are very small. Moreover, (\ref{1condicionesEq}) implies that the movable mirror has both a velocity and an acceleration sufficiently small so that the field evolves as if the movable mirror were instantaneously fixed at its position $q(t)$. In other words, the field \textit{sees} the movable mirror as if it were fixed at $q(t)$.
  
We now proceed to solve (\ref{1}) and (\ref{2}). In order to do this we first introduce the modes of the system for fixed $q(t)$.


\subsection{Modes for fixed q(t)}

For $q(t)$ fixed, the modes associated with (\ref{1}), (\ref{4}), and (\ref{1delta}) were calculated in \cite{Nussenzveig,N2}. Adapting them to our coordinate system one finds that they are given by
\begin{eqnarray}
A_{0,k}(x,t) &=&  V_{k}[x,q(t)] e^{- i\omega t} \ , 
\end{eqnarray}
with
\begin{eqnarray}
\label{7Modos}
&& V_{k}[x,q(t)] 
\ = \
\begin{cases}
L_{k}[q(t)]\mbox{sin}(kx)  \cr
\qquad\qquad  \mbox{if} \ \ \ 0\leq x \leq q(t) \ , \cr
\sqrt{\frac{2}{\pi}}\mbox{sin}\left\{ k\left[ x-q(t) \right] +\delta_{k}\left[ q(t) \right]  \right\} \cr
\qquad\qquad \mbox{if} \ \ \ x > q(t) \ .
\end{cases}
\end{eqnarray}
Here \ $k>0$, \ $\omega = ck$, and
\begin{eqnarray}
\label{7Lk}
L_{k}\left[ q(t) \right] &=& \sqrt{\frac{2}{\pi}} \left\{ \ 1 + (4\pi \chi_{0} k)^{2}\mbox{sin}^{2}\left[ kq(t) \right] \right. \cr
&& \qquad \ \ \left. - (4\pi \chi_{0}k)\mbox{sin}\left[ 2kq(t) \right] \ \right\}^{-1/2} \ , \cr
&& \cr
\mbox{sin}\left\{ \delta_{k}\left[ q(t)  \right] \right\} &=& \sqrt{\frac{\pi}{2}}L_{k}\left[ q(t) \right]\mbox{sin}\left[ kq(t) \right] \ , \cr
&& \cr
\mbox{cos}\left\{ \delta_{k}\left[ q(t)  \right] \right\} &=& \sqrt{\frac{\pi}{2}}L_{k}\left[ q(t) \right]\times\cr
&& \times \left\{ \ \mbox{cos}\left[ kq(t) \right] - (4\pi\chi_{0}k)\mbox{sin}\left[ kq(t) \right] \ \right\} \ . \cr
&&
\end{eqnarray}
Notice that $V_{k}[x,q(t)]$ is a real-valued function. 

We want to use the modes in (\ref{7Modos}) to describe the plane wave of a monochromatic laser approaching the cavity from the right. In order to do this we have to eliminate the mirror-position dependent phase \ $e^{-i\delta_{k}[q(t)]}e^{ikq(t)}$ \ associated with $e^{-ikx}$ in the second line of (\ref{7Modos}). This is done by introducing a phase shift as follows: 
\begin{eqnarray}
\label{Vkt}
&& \tilde{V}_{k}[x,q(t)] \ = \ e^{i\Phi_{k}[q(t)]} V_{k}[x,q(t)] \ , \cr 
&& \cr
&=&
\begin{cases}
\frac{i}{2}L_{k}[q(t)]e^{i\left\{ \delta_{k}[q(t)] - kq(t) \right\}}\left( e^{-ikx} -e^{ikx} \right) \cr
\qquad\qquad \mbox{if} \ \ 0\leq x \leq q(t), \cr
\frac{i}{2}\sqrt{\frac{2}{\pi}}\left\{ e^{-ikx} -e^{ikx}e^{i2\left\{ \delta_{k}[q(t)] - kq(t) \right\} } \right\} \cr
\qquad\qquad \mbox{if} \ \ x > q(t) .
\end{cases}
\cr
&&
\end{eqnarray}
with
\begin{eqnarray}
\label{VktF}
\Phi_{k}[q(t)] \ = \ \delta_{k}[q(t)] -kq(t) \ .
\end{eqnarray}
Now the plane wave $e^{-ikx}$ in the second line of (\ref{Vkt}) can be associated with a monochromatic laser on the far right, since its phase does not depend on the position of the movable mirror. Also, (\ref{Vkt}) expresses the modes in terms of incoming and outgoing waves. Up to a normalization factor, (\ref{Vkt}) correctly describes an incoming plane wave with amplitude $1$ and a reflected plane wave with amplitude given by the scattering matrix $S(k) = e^{i2\left\{ \delta_{k}[q(t)] -kq(t)\right\} }$. The factor $i$ in (\ref{Vkt}) is present so that (\ref{Vkt}) reduces to $\tilde{V}_{k}[x,q(t)] = (2/\pi)^{1/2}\mbox{sin}(kx)$ \ when \ $\chi_{0} = 0$ \ (the movable mirror is completely transparent). Also, (\ref{Vkt}) is $\sqrt{2/\pi}$ times the standard physical solution in scattering in the half-line $(0,+\infty)$ with a Dirichlet boundary condition at \ $x=0$. The factor $\sqrt{2/\pi}$ is necessary for the orthogonality relation presented further below.

For fixed $k$ the transmissivity $T$ of the movable mirror is given by \cite{Nussenzveig}
\begin{eqnarray}
\label{NuevoTransparencia}
T &=& \left[ 1 + \left(\frac{4\pi\chi_{0}k}{2}\right)^{2} \right]^{-1} \ .
\end{eqnarray} 
Hence, the transparency of the movable mirror will be small if \ $T << 1$ \ or, equivalently, \ $4\pi\chi_{0}k \gg 1$. 

Before proceeding we give some properties of $L_{k}\left[ q(t) \right]$ that are deduced in Appendix I. For fixed $k>0$, the function $L_{k}\left[ q(t) \right]$  is maximized (minimized) for a discrete set of values $q_{2n}$ ($q_{2n+1}$) of $q(t)$ with \ $n\in\mathbb{Z}^{+}$. Here $\mathbb{Z}^{+}$ is the set of non-negative integers. If \ $4\pi\chi_{0}k \gtrsim 5$, then one has to good approximation 
\begin{eqnarray}
\label{ExtraMaximizadoresQ2n}
kq_{2n} &\simeq& n\pi  + \frac{1}{4\pi\chi_{0}k} \ , \cr
&& \cr
kq_{2n+1} &\simeq& \left( n + \frac{1}{2} \right)\pi  + \frac{1}{4\pi\chi_{0}k} \ ,
\end{eqnarray}
and
\begin{eqnarray}
\label{ExtraMaximizadoresL}
L_{k}(q_{2n}) &\simeq& \sqrt{\frac{2}{\pi}}(4\pi\chi_{0}k) \ , \cr
&& \cr
L_{k}(q_{2n+1}) &\simeq& \sqrt{\frac{2}{\pi}}\left(\frac{1}{4\pi\chi_{0}k}\right) \ .
\end{eqnarray}
Notice that \ $kq_{2n} \rightarrow n\pi$ \ as \ $4\pi\chi_{0}k \rightarrow +\infty$, that is, $q_{2n}$ tends to the values $n\pi/k$ corresponding to the case where the movable mirror is perfect \cite{Jackson} as the transparency tends to zero.

Also, one can approximate $L_{k}(q)^{2}$ by a Lorentzian if the transparency of the movable mirror is small and one restricts $q$ to an interval around $q_{2n}$ whose endpoints are not near the minimizers $q_{2n \pm 1}$. Explicitly,
\begin{eqnarray}
\label{ExtraLorentzianaL}
L_{k}(q)^{2} &\simeq& \left(\frac{2}{\pi\xi^{2}}\right)\frac{1}{k^{2}(q-q_{2n})^{2} +\frac{1}{\xi^{4}}} \  ,
\end{eqnarray}
with \ $\xi = 4\pi \chi_{0}k$ \ if \ $k\vert q-q_{2n} \vert \ll 3/2$ \ and \ $2/\xi \ll 1$.

\begin{figure}[htbp]
  \centering
  \includegraphics[scale=0.75]{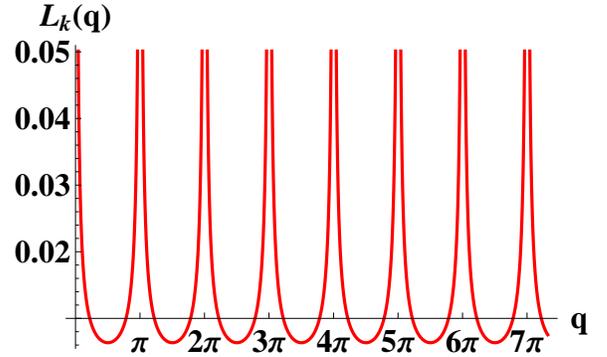}
  \caption{(Color online) The figure illustrates $L_{k}(q)$ as a function of $q$ for $\chi_{0} = 10$ m and $k = 1$ m$^{-1}$. For a better illustration only part of the $y$-axis is shown.}
\label{Figure1}
\end{figure}

We now discuss the behavior of $L_{k}\left[ q(t) \right]$ and its relation to the cavity resonance frequencies in the case where the transparency of the movable mirror is small, that is, in the case \ $4\pi \chi_{0} k \gg 1$. Assume that \ $k>0$ \ is fixed.  As the mirror moves, $L_{k}\left[ q(t) \right]$ will be very large only when \ $q(t) \simeq q_{2n}$ \ for some \ $n\in\mathbb{Z}^{+}$, see (\ref{ExtraMaximizadoresL}) and (\ref{ExtraLorentzianaL}). As a result, \ $\omega = ck$ \ will coincide with one of the cavity resonance frequencies only when the mirror is sufficiently close to one of these special positions. In fact, from (\ref{ExtraLorentzianaL}) it follows that \ $\omega = ck$ \ coincides with one of the cavity resonance frequencies if and only if \ $k\vert q-q_{2n} \vert \leq (4\pi\chi_{0}k)^{-2}$. In other words, the resonant positions $kq_{2n}$ have a half-width-at-half-maximum equal to $(4\pi\chi_{0}k)^{-2}$.  Figure~\ref{Figure1} shows $L_{k}(q)$ as a function of $q$ for $\chi_{0} = 10$ m and $k=1$ m$^{-1}$. 

The modes (\ref{7Modos}) form a continuous set and satisfy orthonormalization and completeness relations \cite{Nussenzveig,N2}. Adapting them to the modes $\tilde{V}_{k}[x,q(t)]$ it follows that
\begin{eqnarray}
\label{7ortonormalizacion}
\delta(k-k') = \int_{0}^{+\infty}dx \ \epsilon\left[ x-q(t) \right] \tilde{V}_{k}\left[ x,q(t) \right]^{*} \tilde{V}_{k'}\left[ x,q(t) \right] , \cr
&&
\end{eqnarray}
and any function \ $f(x)\in \mathcal{L}^{2}\left[0, +\infty \right)$ \ can be expanded in the form 
\begin{eqnarray}
\label{7Completez}
f(x) &=& \int_{0}^{+\infty}dk \ f_{k}\left[ q(t) \right]\tilde{V}_{k}\left[ x,q(t) \right] \ , \
\end{eqnarray}
with the $k$th-mode $f_{k}\left[ q(t) \right]$ given by
\begin{eqnarray}
\label{7CoefFourier}
f_{k}\left[ q(t) \right] &=& \int_{0}^{+\infty}dx \ \epsilon\left[ x-q(t) \right] \tilde{V}_{k}\left[ x,q(t) \right]^{*}f(x) \ . \ \
\end{eqnarray}
Using (\ref{7Completez}) and (\ref{7CoefFourier}) it follows that
\begin{eqnarray}
\label{9}
A_{0}(x,t) &=& \int_{0}^{+\infty} dk \ Q_{k}\left[ t,q(t) \right] \tilde{V}_{k}\left[ x,q(t) \right] \ , 
\end{eqnarray}
with
\begin{eqnarray}
\label{9bis}
Q_{k}\left[t,q(t)\right] &=& \int_{0}^{+\infty}dx \ \epsilon\left[ x-q(t) \right] \tilde{V}_{k}\left[ x,q(t) \right]^{*} A_{0}(x,t) \ . \cr
&&
\end{eqnarray}

Substituting (\ref{9}) into (\ref{1}), neglecting terms proportional to $\dot{q}(t)/c$, $\ddot{q}(t)/(c\omega_{0})$, and higher powers of them ($\omega_{0}$ the characteristic frequency of the field), and using the orthonormalization relation in (\ref{7ortonormalizacion}) one obtains harmonic oscillator equations for each of the modes $Q_{k}\left[ t,q(t) \right]$:
\begin{eqnarray}
\label{25}
0 &=& \frac{d^{2}}{dt^{2}}Q_{k}\left[ t,q(t) \right] + \omega_{k}^{2}Q_{k}\left[ t,q(t) \right] \ ,
\end{eqnarray}
with \ $\omega_{k} = ck$, \ $k>0$, and \ $t\in\mathbb{R}$. We have dropped terms proportional to $\dot{q}(t)/c$, $\ddot{q}(t)/(c\omega_{0})$, and higher powers of them because terms of these orders were neglected to obtain (\ref{1}). Also notice that only by dropping these terms does one recover the physical situation described by (\ref{1}) in which the field evolves as if the movable mirror were fixed.

From (\ref{25}) one immediately obtains for \ $k>0$ \ and \ $t\in\mathbb{R}$ \ that
\begin{eqnarray}
\label{55}
Q_{k}\left[ t,q(t) \right] &=& g(k)e^{-i\omega_{k}t} + g(k)^{*}e^{i\omega_{k}t}e^{-i2\Phi_{k}[q(t)]} \ . \ \ \
\end{eqnarray}
Here we used the fact that $A_{0}(x,t)$ must be a real quantity.


\section{SINGLE-MODE FIELD}

In the rest of the article we assume that the field has a single-mode, that is,
\begin{eqnarray}
\label{70}
g(k) &=& g_{0}e^{i\phi_{0}}\delta\left( k - k_{\scriptscriptstyle{N}}^{\scriptscriptstyle{0}} \right) \ ,
\end{eqnarray}
with \ $k_{\scriptscriptstyle{N}}^{\scriptscriptstyle{0}}, g_{0} >0$ \ and \ $\phi_{0} \in \mathbb{R}$. 

We now discuss which physical situation is described approximately by (\ref{70}). A monochromatic laser on the far right is always turned on. The plane wave associated with the laser travels to the left, is partially reflected by the movable mirror at \ $q(t)>0$, and completely reflected by the perfect mirror fixed at \ $x=0$. After a transient, a standing wave is approximately formed and the laser is responsible for maintaining it. This standing wave is defined by the mode in (\ref{Vkt}) with \ $k = k_{\mbox{\tiny N}}^{\mbox{\tiny 0}}$. We describe the dynamics of the system after the aforementioned transient. Notice that the electromagnetic field does not decay with time because the laser is always turned on. Moreover, the field inside the cavity does not decay irreversibly with time because the laser is always driving it. It is important to note that the restriction to a single-mode is possible when terms of order $\dot{q}(t)/c$ are neglected (which is our case). If such terms are maintained, then field amplitudes corresponding to different wave-numbers are mixed \cite{Nuestro,DomokosI}. Recall that we are considering the case where \ $\vert \dot{q}(t)/c \vert \ll 1$ so that this process is very small. 

Substituting (\ref{55}) and (\ref{70}) in  (\ref{9}) and choosing the origin of time so that \ $\phi_{0} = 0$ \  it follows that
\begin{eqnarray}
\label{72}
A_{0}(x,t) &=& 2 g_{0}\mbox{cos}\left( \omega_{0}t + k_{\scriptscriptstyle{N}}^{\scriptscriptstyle{0}}q(t)  -\delta_{k_{\scriptscriptstyle{N}}^{\scriptscriptstyle{0}}}[q(t)] \right) \times \cr
&& \qquad \times V_{k_{N}^{0}}\left[x,q(t)\right] \ , 
\end{eqnarray}
 with \ $\omega_{0} \equiv \omega_{k_{\scriptscriptstyle{N}}^{\scriptscriptstyle{0}}} = c k_{\scriptscriptstyle{N}}^{\scriptscriptstyle{0}}$. Using the complex-valued modes $\tilde{V}_{k}[x,q(t)]$ in (\ref{Vkt}) one can also express $A_{0}(x,t)$ in the form
\begin{eqnarray}
\label{FormaCompleja}
A_{0}(x,t) &=& \mbox{Re}\left\{ \ 2g_{0}\tilde{V}_{k_{\scriptscriptstyle{N}}^{0}}[x,q(t)]e^{-i\omega_{0}t}  \ \right\} \ .
\end{eqnarray}
Here Re is the real part of a complex number.
 
All that remains is to solve equation (\ref{2}) for the moving mirror. We first express (\ref{2}) in terms of non-dimensional quantities.

Define
\begin{eqnarray}
\label{76}
\Delta &=& \sqrt{ \frac{g_{0}^{2}\omega_{0}^{3}}{\pi^{2}M_{0}c^{3}}} \ , \cr
&& \cr
\xi &=& 4\pi\chi_{0}k_{\scriptscriptstyle{N}}^{\scriptscriptstyle{0}} \ , \cr
&& \cr
\Omega &=& 2\frac{\omega_{0}}{\Delta} \ , \cr
&& \cr
x(\tau) &=& k_{\scriptscriptstyle{N}}^{\scriptscriptstyle{0}}q\left( \frac{\tau}{\Delta} \right) \ .
\end{eqnarray}
Here $\Delta$ has units of 1/s, $\tau = \Delta t$ is the non-dimensional time, and $\xi$, $\Omega$, and $x(\tau)$ are non-dimensional quantities. Notice that have taken $1/k_{\scriptscriptstyle{N}}^{\scriptscriptstyle{0}}$ to be the characteristic length of the system, while $1/\Delta$ is the characteristic time. In other words, we have chosen to measure lengths in units of one over the wave-number of the field, while time is measured in units of a quantity involving the strength $g_{0}$ and angular frequency $\omega_{0}$ of field and the mass per unit area $M_{0}$ of the movable mirror. The form of $\Delta$ was dictated by the differential equation in (\ref{2}) and can be thought of as $1$ over the time-scale in which the position of the movable mirror changes appreciably. Therefore, \ $\Omega = (4\pi/\Delta)/(2\pi/\omega_{0})$ \ is interpreted to be the time-scale $4\pi/\Delta$ in which the movable mirror changes appreciably divided by the time-scale $2\pi/\omega_{0}$ in which the field changes appreciably. Since the field normally evolves on a much smaller time-scale than that of the movable mirror, one should expect $\Omega > 1$. Also, from (\ref{NuevoTransparencia}) and (\ref{76}) it follows that the transmissivity $T$ of the movable mirror is given by 
\begin{eqnarray}
\label{NuevoTransparencia2}
T &=& \left[ 1+ \left( \frac{\xi}{2} \right)^{2} \right]^{-1} \ ,
\end{eqnarray}
so that the transparency will be small if $\xi$ is large.

Using (\ref{72}) and (\ref{76}) it can be shown that (\ref{2}) can be rewritten as follows:
\begin{eqnarray}
\label{80}
\frac{d^{2}x}{d\tau^{2}}(\tau) &=& \left\{ 1+ \mbox{cos}\left[ \Omega\tau +2x(\tau) -2\delta_{k_{\scriptscriptstyle{N}}^{\scriptscriptstyle{0}}}\left( \frac{x(\tau)}{k_{\scriptscriptstyle{N}}^{\scriptscriptstyle{0}}} \right) \right] \right\} \times \cr 
&& \qquad \times f_{\mbox{\tiny RWA}}\left[ x(\tau) \right] \ ,
\end{eqnarray} 
with
\begin{eqnarray}
\label{81}
f_{\mbox{\tiny RWA}}(x) &=& -\frac{1}{2}\left[ 1 - \frac{1}{1 + \xi^{2}\mbox{sin}^{2}(x) - \xi \mbox{sin}(2x)} \right] \ . \ \
\end{eqnarray}
Note that \ $1 + \xi^{2}\mbox{sin}^{2}(x) - \xi \mbox{sin}(2x) >0$ \ for all \ $x\in \mathbb{R}$. The notation $f_{\mbox{\tiny RWA}}(x)$ has been chosen because the term on the right of (\ref{80}) will reduce to $f_{\mbox{\tiny RWA}}[x(\tau)]$ in the rotating wave approximation (RWA). We note that if one expands (\ref{81}) using Taylor series, neglects terms of order $\xi^{n}$ with \ $n\geq 3$, and makes the appropriate changes both in coordinate systems and units, then one obtains the same force as in \cite{DomokosI} for the case in which the transparency of the movable mirror is large, terms of order $\dot{q}(t)/c$ are neglected, and the RWA is performed. 

In this article derivatives with respect to $\tau$ are also denoted by a prime, while derivatives with respect to $t$ are denoted by a dot.

Before proceeding we mention that using (\ref{76}) one can introduce dimensions so that (\ref{80}) takes the form 
\begin{eqnarray}
\label{ExtraDimensiones}
M_{0}\ddot{q}(t) &=& \left( \frac{g_{0}\omega_{0}}{\pi c} \right)^{2}f_{\mbox{\tiny RWA}}\left[ k_{\scriptscriptstyle{N}}^{0}q(t)\right] \times \cr
&& \times \left\{ 1+ \mbox{cos}\left[ 2\omega_{0}t + 2k_{\scriptscriptstyle{N}}^{\scriptscriptstyle{0}}q(t) -2\delta_{k_{\scriptscriptstyle{N}}^{\scriptscriptstyle{0}}}[q(t)] \ \right] \ \right\} \ . \cr
&&
\end{eqnarray}


\subsection{The radiation pressure potential}  

The non-dimensional force $f_{\mbox{\tiny RWA}}(x)$ can be derived from a non-dimensional potential $V_{\mbox{\tiny RWA}}(x)$: 
\begin{eqnarray}
\label{ii4}
f_{\mbox{\tiny RWA}}(x) &=& -\frac{d}{dx}V_{\mbox{\tiny RWA}}(x) \ . \ \ \ \
\end{eqnarray}
If \ $(2m-1)\pi/2 \leq x \leq (2m+1)\pi/2$ \ and \ $x\geq 0$ \ with \ $m \in \mathbb{Z}^{+}$, then $V_{\mbox{\tiny RWA}}(x)$ is given by
\begin{eqnarray}
\label{PotencialRWA}
V_{\mbox{\tiny RWA}}(x) &=& \frac{x}{2} - \frac{1}{2}\mbox{tan}^{-1}\left[ \ (1 + \xi^{2})\mbox{tan}(x) - \xi \ \right]  \cr
&& - \frac{1}{2}\left[ \mbox{tan}^{-1}(\xi) + m\pi \ \right] \ .
\end{eqnarray} 
The expression for $V_{\mbox{\tiny RWA}}(x)$ was obtained by integrating (\ref{81}) from \ $x'=0$ \ to \ $x'=x >0$. Also, tan$^{-1}\theta \in (-\pi/2, \pi/2)$ \ for \ $\theta \in \mathbb{R}$.

We now list properties of $V_{\mbox{\tiny RWA}}(x)$ and $f_{\mbox{\tiny RWA}}(x)$ that will be used throughout the article. They are deduced in Appendix II. Also, figure~\ref{Figure2} illustrates $V_{\mbox{\tiny RWA}}(x)$ for $\xi = 1, 10,$ and $50$.
\begin{enumerate}
\item Periodicity of $V_{\mbox{\tiny RWA}}(x)$: $V_{\mbox{\tiny RWA}}(x)$ is a periodic function of period $\pi$. This follows from the piecewise definition of $V_{\mbox{\tiny RWA}}(x)$ on intervals of length $\pi$ and the fact that $\mbox{tan}(x)$ is periodic of period $\pi$.

\item Maximizers of $V_{\mbox{\tiny RWA}}(x)$: They are denoted by $x_{n}^{*}$ with \ $n\in\mathbb{Z}^{+}$. One has   
\begin{equation} 
\label{MaximizadorVRWA}
x_{n}^{*} \ = \ n\pi \ , \qquad V_{\mbox{\tiny RWA}}(x_{n}^{*}) = 0 \ . 
\end{equation}
Also, $0$ is the absolute maximum value of $V_{\mbox{\tiny RWA}}(x)$.

\item Minimizers of $V_{\mbox{\tiny RWA}}(x)$: They are denoted by $x_{n}^{**}$ with \ $n\in\mathbb{Z}^{+}$. For \ $\xi \gtrsim 5$ \ and \ $n\in\mathbb{Z}^{+}$ \ one has to good approximation   
\begin{eqnarray}
\label{MinimizadorVRWA} 
x_{n}^{**} &\simeq& n\pi + 2/\xi \  , \cr
V_{\mbox{\tiny RWA}}(x_{n}^{**}) &\simeq& V_{\mbox{\tiny RWA}}(2/\xi) \simeq -\mbox{tan}^{-1}(\xi) \ .
\end{eqnarray}
Also, $V_{\mbox{\tiny RWA}}(x_{n}^{**})$ is the absolute minimum value of $V_{\mbox{\tiny RWA}}(x)$. 

\item Bounds of $V_{\mbox{\tiny RWA}}(x)$: For all \ $x \geq 0$ \ one has 
\begin{eqnarray}
\label{CotasVRWA}
-\frac{\pi}{2} \ < \  V_{\mbox{\tiny RWA}}(x) \ \leq \ 0 \ .
\end{eqnarray}
Also, \ $V_{\mbox{\tiny RWA}}(2/\xi) \rightarrow -\pi/2$ \ as \ $\xi \rightarrow + \infty$.

\item Wells of $V_{\mbox{\tiny RWA}}(x)$: They all have non-dimensional length $\pi$, which corresponds to a length of half the wavelength of the field, that is, $\lambda_{0}/2 = \pi/k_{\scriptscriptstyle{N}}^{\scriptscriptstyle{0}}$. Moreover, the depth of the wells increases as the the transparency of the mirror decreases, that is, as \ $\xi \rightarrow +\infty$. 

\item Maximizers of $f_{\mbox{\tiny RWA}}(x)$: They are denoted by $x_{2n}$ \ with $n\in\mathbb{Z}^{+}$. For \ $\xi \gtrsim 5$ \ and \ $n\in\mathbb{Z}^{+}$ \ one has to good approximation 
\begin{equation}
\label{MaximizadorfRWA}
x_{2n} \ \simeq \ n\pi + \frac{1}{\xi} \ , \qquad f_{\mbox{\tiny RWA}}(x_{2n}) \ \simeq \ \frac{\xi^{2}}{2} \ .
\end{equation}

\item Minimizers of $f_{\mbox{\tiny RWA}}(x)$: They are denoted by $x_{2n+1}$ \ with $n\in\mathbb{Z}^{+}$. For \ $\xi \gtrsim 5$ \ and \ $n\in\mathbb{Z}^{+}$ \ one has to good approximation
\begin{equation}
\label{MinimizadorfRWA}
x_{2n+1} \simeq \left( n+\frac{1}{2} \right) \pi + \frac{1}{\xi} \ , \ \ f_{\mbox{\tiny RWA}}(x_{2n+1}) \simeq -\frac{1}{2}   
\end{equation}

\item Zeros of $f_{\mbox{\tiny RWA}}(x)$: They are the maximizers $x_{n}^{*}$ and minimizers $x_{n}^{**}$ of $V_{\mbox{\tiny RWA}}(x)$.

\item Approximation by a Lorentzian: If \ $\xi \gg 1$, \ $x\geq 0$, \ $u = x-(m\pi +1/\xi)$, \ $1 \gg 3^{-1}(u + \xi^{-1})^{2}$, \ and \ $(2m-1)\pi/2 \leq x \leq (2m+1)\pi/2$ \ for some \ $m\in\mathbb{Z}^{+}$, then 
\begin{equation}
\label{ExtraLorentziana}
f_{\mbox{\tiny RWA}}(x) \ \simeq \ -\frac{1}{2} + \frac{1}{2\xi^{2}}\left( \frac{1}{u^{2} + \xi^{-4}} \right) \ .
\end{equation} 
Calculating numerically the relative error between $f_{\mbox{\tiny RWA}}(x)$ and the displaced Lorentzian on the right-hand side of (\ref{ExtraLorentziana}), we found that the displaced Lorentzian approximates well  $f_{\mbox{\tiny RWA}}(x)$ if \ $\xi \gtrsim 10$ \ and \ $-1/\xi < u < 1/\xi$.
\end{enumerate} 

\begin{figure}[htbp]
  \centering
  \includegraphics[scale=0.75]{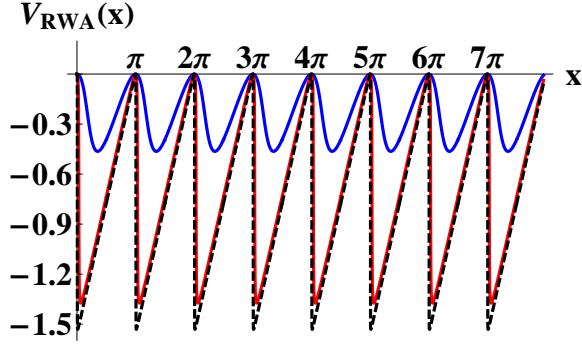}\\
  \caption{(Color online) The figure shows $V_{\mbox{\tiny RWA}}(x)$ for $\xi = 1$ (blue-solid line), $10$ (red-solid line), and $50$ (black-dashed line).}
\label{Figure2}
\end{figure}


\subsection{Approximation for $V_{\mbox{\tiny RWA}}(x)$}

From Fig.~\ref{Figure2} it is clear that $V_{\mbox{\tiny RWA}}(x)$ tends to a \textit{saw tooth wave} as \ $\xi \rightarrow +\infty$. Hence, it can be approximated by a continuous linear interpolating polynomial for sufficiently large $\xi$ (say, \ $\xi \gtrsim 50$), that is, $V_{\mbox{\tiny RWA}}(x)$ can be approximated as follows:
\begin{eqnarray}
\label{91o}
&& V_{\mbox{\tiny RWA}}(x) \ \simeq \ V_{\mbox{\tiny RWA}}^{\mbox{\tiny approx}}(x) \ , 
\end{eqnarray}
with
\begin{eqnarray}
\label{91}
V_{\mbox{\tiny RWA}}^{\mbox{\tiny approx}}(x) 
&\equiv& 
\begin{cases}
0 \qquad\ \ \ \ \mbox{if} \ \ \ n\pi \leq x \leq n\pi +\frac{1}{\xi} \ , \cr
V_{\mbox{\tiny RWA}}\left(\frac{2}{\xi} \right)  \cr
\qquad\qquad \mbox{if} \ \ n\pi +\frac{1}{\xi} < x \leq n\pi + \frac{2}{\xi} \ , \cr
V_{\mbox{\tiny RWA}}\left(\frac{2}{\xi} \right) + m_{+}\left[ \ x- \left( n\pi + \frac{2}{\xi} \right) \ \right] \cr
\qquad\qquad \mbox{if} \ \ n\pi +\frac{2}{\xi} \leq x \leq (n+1)\pi  \  .
\end{cases}
\cr
&&
\end{eqnarray}
for \ $n\pi \leq x \leq (n+1)\pi$, \ $n\in\mathbb{Z}^{+}$. Here
\begin{eqnarray}
\label{91bis}
m_{+} &=& -\frac{V_{\mbox{\tiny RWA}}(2/\xi)}{\pi -2/\xi} \ \simeq \ \frac{1}{2} \ .
\end{eqnarray}
We note that the point \ $x = (n\pi +1/\xi)$ \ corresponds (approximately) to a maximizer $x_{2n}$ of $f_{\mbox{\tiny RWA}}(x)$, see (\ref{MaximizadorfRWA}).

\begin{figure}[htbp]
  \centering
  \subfloat[]{\label{Figure3a}\includegraphics[scale=0.75]{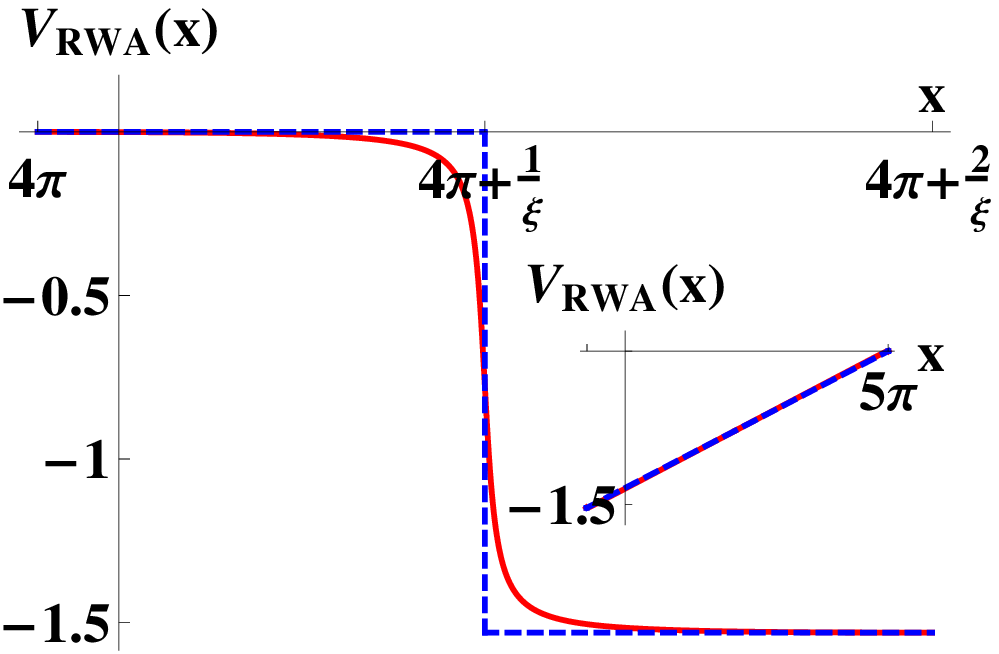}} \hspace{0.3cm}
  \subfloat[]{\label{Figure3b}\includegraphics[scale=0.75]{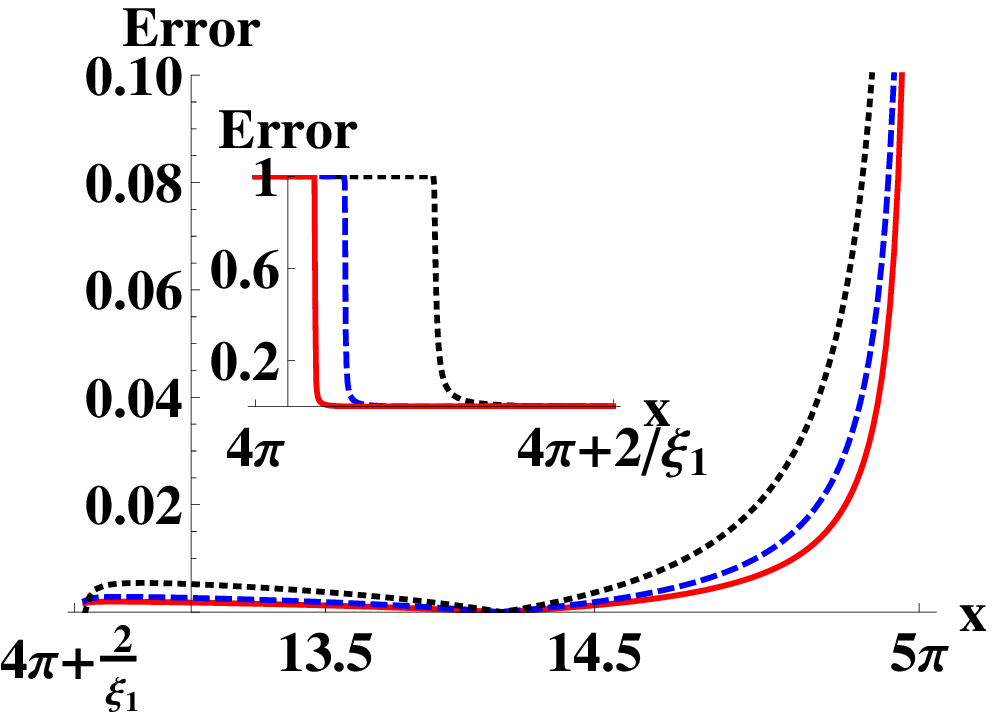}}\\
  \caption{
(Color online) Figure~\ref{Figure3a} shows a close up of one of the wells of $V_{\mbox{\tiny RWA}}(x)$ (red-solid line) and the approximate $V_{\mbox{\tiny RWA}}^{\mbox{\tiny approx}}(x)$ (blue-dashed line) for \ $\xi = 50$ \ and the region \ $[4\pi,4\pi +2\xi^{-1}]$. The inside figure shows the region \ $[4\pi +2\xi^{-1},5\pi]$. Figure~\ref{Figure3b} illustrates the relative error \ $\vert  V_{\mbox{\tiny RWA}}(x) - V_{\mbox{\tiny RWA}}^{\mbox{\tiny approx}}(x)\vert / \vert  V_{\mbox{\tiny RWA}}(x) \vert$ \ for $\xi_{1} = 50$ (black-dotted line), $\xi_{2} = 100$ (blue-dashed line), and $\xi_{3} = 150$ (red-solid line). The inside figure shows a close-up of the interval $[4\pi , 4\pi +2/\xi_{1}]$.}
\label{Figure3}
\end{figure}

Figure~\ref{Figure3a} compares the exact $V_{\mbox{\tiny RWA}}(x)$ with the approximate $V_{\mbox{\tiny RWA}}^{\mbox{\tiny approx}}(x)$ for one of the potential wells and \ $\xi = 50$. Also, figure~\ref{Figure3b} illustrates the relative error \ $\vert  V_{\mbox{\tiny RWA}}(x) - V_{\mbox{\tiny RWA}}^{\mbox{\tiny approx}}(x)\vert / \vert  V_{\mbox{\tiny RWA}}(x) \vert$ \ for $\xi_{1} = 50$ (black-dotted line), $\xi_{2} = 100$ (blue-dashed line), $\xi_{3} = 150$ (red-solid line). From the figures it is clear that the approximation is very good except at thin layers at \ $x=4\pi$, $x= (4\pi +1/\xi)$, and \ $x= 5\pi$. The reason for this is that $V_{\mbox{\tiny RWA}}^{\mbox{\tiny approx}}(x)$ does not take into account the curvature of the potential. Nevertheless, the region over which  $V_{\mbox{\tiny RWA}}^{\mbox{\tiny approx}}(x)$ is a good approximation of $ V_{\mbox{\tiny RWA}}(x)$ increases as \ $\xi \rightarrow + \infty$.

From (\ref{ii4}) and (\ref{91}) it follows that
\begin{eqnarray}
\label{93}
f_{\mbox{\tiny RWA}}(x) &\simeq&  
\begin{cases}
0 & \mbox{if} \ \ n\pi < x < \left( n\pi + \frac{2}{\xi} \right) \ \mbox{and} \cr
& \qquad  x \not=  \left( n\pi + \frac{1}{\xi} \right) \ , \cr
-m_{+} & \mbox{if} \ \ \left( n\pi + \frac{2}{\xi} \right) < x < (n+1)\pi \ .
\end{cases} \cr
&&
\end{eqnarray}
Hence, for large $\xi$ the function $f_{\mbox{\tiny RWA}}(x)$ is approximately a (non-dimensional) piecewise constant force whose magnitude and direction depends on whether the mirror is to the right or to the left of the minimizer \ $x_{n}^{**} \simeq n\pi + 2/\xi$ \ of the potential $V_{\mbox{\tiny RWA}}(x)$.

\section{DYNAMICS DUE TO RADIATION PRESSURE}

Equation (\ref{80}) is a second order non-linear and non-autonomous differential equation. In the following three subsections we will treat it in various regimes.

\subsection{The limit of high field intensity}

The limit of high field intensity is defined by the condition $\Omega \ll 1$. The name \textit{high field intensity} comes from the fact that $\Omega$ is inversely proportional to $g_{0}$. Since \ $\Omega = 2\omega_{0}/\Delta$, it follows that the characteristic time-scale $1/\Delta$ for the evolution of the movable mirror is much smaller than the characteristic time-scale $2\pi/\omega_{0}$ for the evolution of the field. Therefore, this limit is incompatible with equations (\ref{1}) and (\ref{2}) used in the model (recall that these equations require that the field sees the mirror as instantaneously fixed). Consequently, we do not consider this limit in the rest of the article.

To end this subsection we derive a condition that explicitly states that the limit of high field intensity is incompatible with the model used. Consider the second part of (\ref{1condicionesEq}). From (\ref{80}) one finds that
\begin{eqnarray}
\label{ValidezCampov}
\vert x''(\tau) \vert &\leq& 2f_{\mbox{\tiny RWA}}(x_{2n}) \qquad (\tau \in \mathbb{R}),
\end{eqnarray}
where $x_{2n}$ is a maximizer of $f_{\mbox{\tiny RWA}}(x)$, see (\ref{MaximizadorfRWA}). Since \ $\ddot{q}(t)/(c\omega_{0}) = 4x''(\tau)/\Omega^{2}$ \ with \ $\tau = \Delta t$, it follows from (\ref{ValidezCampov}) that
\begin{eqnarray}
\label{ValidezCampovi}
8f_{\mbox{\tiny RWA}}(x_{2n}) \ll \Omega^{2} &\Rightarrow& \left\vert \frac{\ddot{q}(t)}{c \omega_{0}} \right\vert \ \ll \ 1 \ \ \ (t\in\mathbb{R}).
\end{eqnarray}
It is important to note that \ $8f_{\mbox{\tiny RWA}}(x_{2n}) \ll \Omega^{2}$ \ essentially says that $\Omega \gg 1$. For example, if \ $\xi \gtrsim 5$, then the left-hand side of (\ref{ValidezCampovi}) takes the form \ $4\xi^{2} \ll \Omega^{2}$ \ (see (\ref{MaximizadorfRWA})).
 
Since the sufficient condition in (\ref{ValidezCampovi}) needs  \ $\Omega \gg 1$ and the high field limit condition requires \ $\Omega \ll 1 $, we conclude that the high field limit is generally incompatible with the model presented in the first section. This result is reasonable, since a very strong field could subject the mirror to large accelerations.

\subsection{The limit of low field intensity}

The limit of low field intensity or RWA regime is defined by the condition 
\begin{eqnarray}
\label{CondicionIntensidadDebil}
\frac{2\pi}{\Omega} &\ll& P \ ,
\end{eqnarray}
where \ $P\sim 1$ \ is the (non-dimensional) time scale in which $x(\tau)$ changes appreciably. The name \textit{low field intensity} comes from the fact that $2\pi/\Omega$ is proportional to $g_{0}$, while the name \textit{RWA regime} comes from the fact that (\ref{CondicionIntensidadDebil}) needs to be satisfied in order to be able to perform the RWA. 

In the RWA regime the cosine term in (\ref{80}) oscillates very rapidly and averages to zero. Hence, one can perform the RWA to obtain the approximate equation
\begin{eqnarray}
\label{N1RWA}
\frac{d^{2}x}{d\tau^{2}}(\tau) &=& f_{\mbox{\tiny RWA}}\left[ x(\tau) \right] \ . \
\end{eqnarray}
The RWA regime appears to be a natural setting for the model under study, since one requires \ $\Omega \gg 1$ \ and a typical value is \ $\Omega \geq 10^{9}$ (see Appendix III). The large value of \ $\Omega = 2\omega_{0}/\Delta$ \ comes from the fact that there are two separate time scales: the characteristic time for the evolution of the field \ $2\pi/\omega_{0}$ \ which is much smaller than the characteristic time for the evolution of the mirror \ $1/\Delta$.

\begin{figure}[htbp]
  \centering
  \includegraphics[scale=0.75]{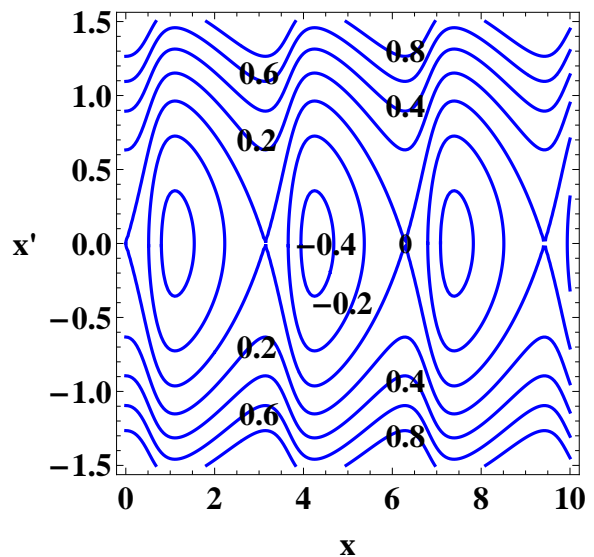}\\
  \caption{(Color online) The figure shows a contour plot of the (non-dimensional) energy $E(x,x')$ of the movable mirror as a function of $x$ and $x'$ for $\xi = 1$. The contours \ $E(x,x') =$ $\pm 0.4$, $\pm 0.2$, $0$, $0.6$, $0.8$ \ are shown.}
\label{Figure4}
\end{figure}

We now analyze and solve to good approximation (\ref{N1RWA}). From (\ref{N1RWA}) one finds that the (non-dimensional) energy $E\left[ x(\tau), x'(\tau) \right]$ of the movable mirror is conserved and is given by
\begin{eqnarray}
\label{N14}
E\left[ x(\tau), x'(\tau) \right] &=& \frac{1}{2}\left[ x'(\tau) \right]^{2} + V_{\mbox{\tiny RWA}}\left[ x(\tau) \right] \ .
\end{eqnarray}
Figure~\ref{Figure4} shows a contour plot of the energy as a function of $x$ and $x'$ for \ $\xi = 1$. 

Since the mirror moves under the potential $V_{\mbox{\tiny RWA}}(x)$ and energy is conserved, it follows that its motion is bounded to one of the wells if and only if \ $E(x,x') \leq 0$. In this case the movable mirror has a periodic trajectory and acquires its maximum speed $v_{0}$ when it is located at one of the minimizers $x_{n}^{**}$ of the potential $V_{\mbox{\tiny RWA}}(x)$. Moreover, the requirement that  \ $E(x,x') \leq 0$ \ is equivalent to the following condition on $v_{0}$:
\begin{eqnarray}
\label{N14v0}
v_{0} &\leq& \sqrt{2\left\vert V_{\mbox{\tiny RWA}}(x_{n}^{**})  \right\vert} \ \leq \ \sqrt{\pi } .
\end{eqnarray} 
In the last inequality we used (\ref{CotasVRWA}). 

Also, the maximizers \ $x_{n}^{*} = n \pi$ \ of $V_{\mbox{\tiny RWA}}(x)$, see (\ref{MaximizadorVRWA}), are points of unstable equilibrium, since a linear stability analysis of the first order system equivalent to (\ref{N1RWA}) shows that \ $(x,x') = (n \pi, 0)$ \ are saddle points (See Theorem 3 in Section 2.10 of \cite{Strogatz}). 
 
Recall that a fixed point (or equilibrium point or critical point) of the first order system equivalent to (\ref{N1RWA}) is a saddle point if there exist two phase space trajectories that approach the fixed point as $\tau \rightarrow +\infty$ and two phase space trajectories that approach the fixed point as $\tau \rightarrow -\infty$ and if there is a neighborhood $V$ of the fixed point such that all other phase space trajectories which start in the deleted neighborhood associated with $V$ leave $V$ as $\tau \rightarrow \pm \infty$. Here a deleted neighborhood of a fixed point is a neighborhood of the fixed point that does not contain the fixed point. 

\begin{figure}[htbp]
  \centering
  \subfloat[]{\label{Figure5a}\includegraphics[scale=0.55]{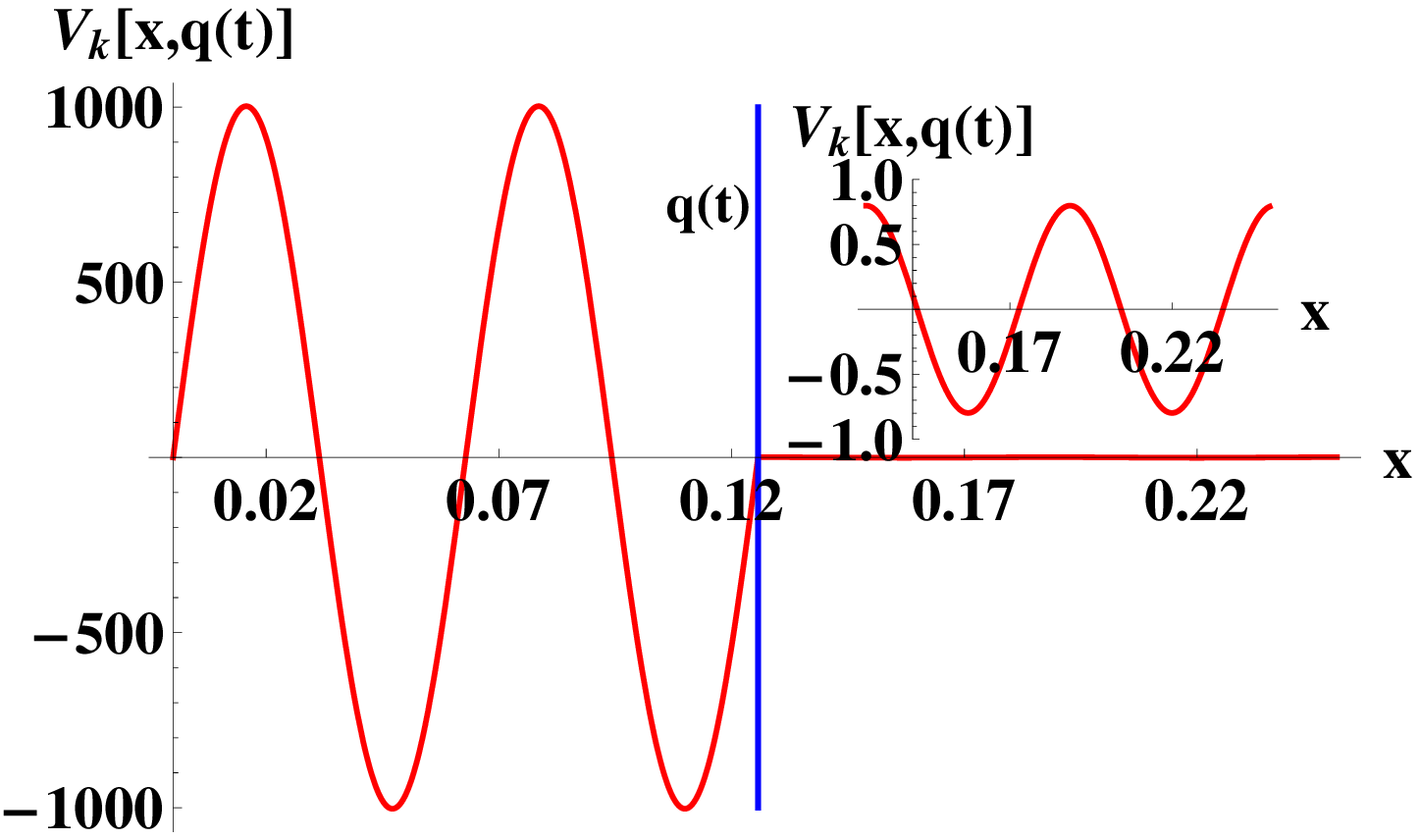}} \hspace{0.3cm}
  \subfloat[]{\label{Figure5b}\includegraphics[scale=0.55]{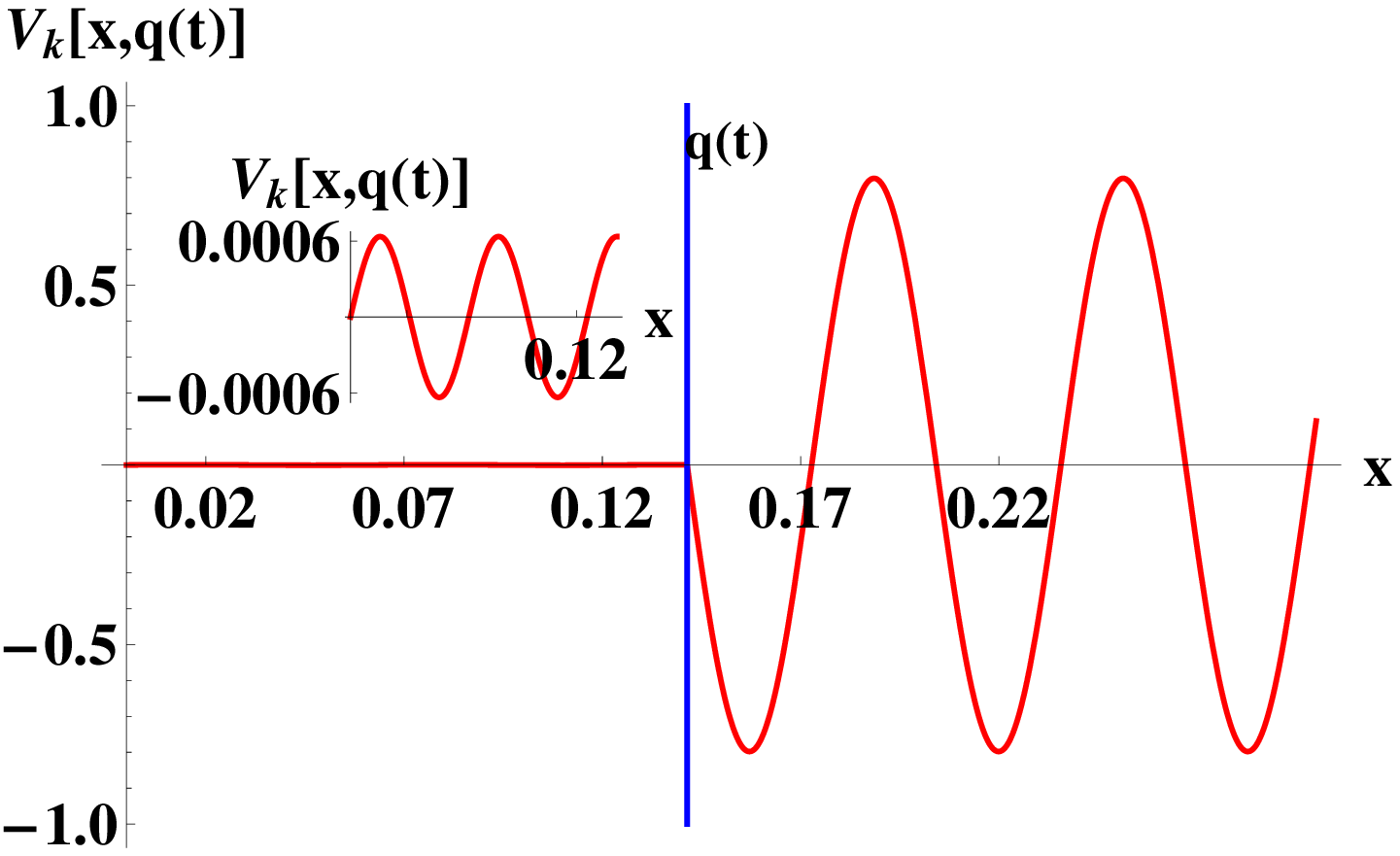}} \\
  \caption{(Color online) The figures illustrate $V_{k_{\scriptscriptstyle{N}}^{\scriptscriptstyle{0}}}[x,q(t)]$ as a function of $x$ when the movable mirror is located at the position $q(t)$ with $\chi_{0} = 1$ m, $k_{\scriptscriptstyle{N}}^{\scriptscriptstyle{0}} = 100$ m$^{-1}$, $n=4$, and \ $\xi = 4\pi\chi_{0}k_{\scriptscriptstyle{N}}^{0} = 4\pi\times 100$. In all figures the vertical blue line shows the position of the movable mirror. In Fig.~\ref{Figure5a} \ $k_{\scriptscriptstyle{N}}^{\scriptscriptstyle{0}}q(t) = n\pi + 1/\xi$ \ and $\omega_{0}$ coincides with one of the cavity resonance frequencies. The inside figure shows $V_{k_{\scriptscriptstyle{N}}^{\scriptscriptstyle{0}}}[x,q(t)]$ outside of the cavity. In Fig.~\ref{Figure5b} \ $k_{\scriptscriptstyle{N}}^{\scriptscriptstyle{0}}q(t) = (n + 1/2)\pi$ \ is such that $\omega_{0}$ is very different from all cavity resonance frequencies. The inside figure shows $V_{k_{\scriptscriptstyle{N}}^{\scriptscriptstyle{0}}}[x,q(t)]$ inside the cavity.}
\label{Figure5}
\end{figure}

One can understand the bounded periodic motion of the movable mirror by analyzing the behavior of the mode $V_{k_{\scriptscriptstyle{N}}^{\scriptscriptstyle{0}}}[x,q(t)]$ of $A_{0}(x,t)$ in (\ref{72}). Assume that the mirror is initially located at a position \ $q(t) = q_{2n}$ \ such that $\omega_{0}$ coincides with one of the cavity resonance frequencies, see (\ref{ExtraMaximizadoresQ2n}). Then $V_{k_{\scriptscriptstyle{N}}^{\scriptscriptstyle{0}}}[x,q(t)]$ will be very large inside the cavity and small outside of it, see Fig.~\ref{Figure5a}. As a result, the mirror is pushed to the right. As the mirror moves to the right, $\omega_{0}$ will become very different from one of the cavity resonance frequencies and $V_{k_{\scriptscriptstyle{N}}^{\scriptscriptstyle{0}}}[x,q(t)]$ will decrease inside the cavity. At some point $V_{k_{\scriptscriptstyle{N}}^{\scriptscriptstyle{0}}}[x,q(t)]$ will be smaller inside the cavity than outside of it, see Fig.~\ref{Figure5b}. As a result, the mirror will slow down and eventually start moving to the left. Then the mirror will approach the \textit{resonant position} $q_{2n}$ and, as a result, $V_{k_{\scriptscriptstyle{N}}^{\scriptscriptstyle{0}}}[x,q(t)]$ will increase inside the cavity. At some point it will be larger inside the cavity than outside of it, see Fig.~\ref{Figure5a}. The mirror will then slow down and eventually start moving to the right again. This process repeats itself over and over giving rise to the periodic motion. Therefore, the bounded motion of the mirror is determined by the mirror approaching or withdrawing from a position where the frequency of the field (approximately) coincides with one of the cavity resonance frequencies. Notice that in the case of bounded motion and \ $\xi \gtrsim 5$, for each potential well there is only one position of the mirror, namely \ $kq_{2n} = x_{2n} \simeq n\pi + 1/\xi$ \ for some \ $n\in\mathbb{Z}^{+}$, such that the frequency $\omega_{0}$ of the field coincides with one of the cavity resonance frequencies.

The unbounded motion of the mirror is understood by referring to the potential $V_{\mbox{\tiny RWA}}(x)$. If the movable mirror has positive energy (see Fig.~\ref{Figure4}), then it accelerates (decelerates) as it approaches a minimizer (maximizer) of $V_{\mbox{\tiny RWA}}(x)$. In this case the radiation pressure is never sufficiently large so as to decelerate the mirror to a complete stop as in the case of the bounded motion.

Now assume that \ $\xi \gg 1$ \ (say, \ $\xi \gtrsim 50$), that is, the mirror has small transparency. In this case one can use the approximation in (\ref{91}) for $V_{\mbox{\tiny RWA}}(x)$ to solve analytically equation (\ref{N1RWA}). We shall only consider the case in which \ $E( x,x' ) < 0$. Notice that one cannot describe accurately the bounded motion of the mirror in a small band around \ $E( x,x' ) = 0$, since the mirror cannot approach \ $x=n\pi$ with the approximation in (\ref{91}).

Assuming that the moving mirror is in the $n$-th well and pasting the solutions one gets that
\begin{eqnarray}
\label{N15}
&& x_{\mbox{\tiny RWA}}(\tau) \cr
&=& 
\begin{cases}
\left( n\pi + \frac{2}{\xi} \right) + v_{0}(\tau - \tau_{\scriptscriptstyle{2k}}) -\frac{m_{+}}{2}(\tau -\tau_{\scriptscriptstyle{2k}})^{2}   \cr
\qquad\qquad\qquad\qquad\qquad \mbox{if} \ \ \tau_{\scriptscriptstyle{2k}} \leq \tau \leq \tau_{\scriptscriptstyle{2k+1}} \ , \cr
\left( n\pi + \frac{2}{\xi} \right) - v_{0}(\tau - \tau_{\scriptscriptstyle{2k+1}}) \cr
\qquad\qquad\qquad\qquad\qquad \mbox{if} \ \ \tau_{\scriptscriptstyle{2k+1}} \leq \tau \leq \tau_{\scriptscriptstyle{2k+1}}' \ , \cr
\left( n\pi + \frac{1}{\xi} \right) + v_{0}(\tau - \tau_{\scriptscriptstyle{2k+1}}') \cr
\qquad\qquad\qquad\qquad\qquad \mbox{if} \ \ \tau_{\scriptscriptstyle{2k+1}}' \leq \tau \leq \tau_{\scriptscriptstyle{2(k+1)}} \ .
\end{cases} \cr
&&
\end{eqnarray}
Here \ $k \in\mathbb{Z}$ \ and $\tau_{0}$ is an instant such that \ $x(\tau_{0}) = n\pi + 2/\xi$ \ and \ $v_{0} = (dx/d\tau)(\tau_{0}) > 0$ \ is the maximum (non-dimensional) speed of the mirror. For \ $k \in \mathbb{Z}$ \ one has 
\begin{eqnarray}
\label{N14tiempos}
\tau_{\scriptscriptstyle{2k+1}} - \tau_{\scriptscriptstyle{2k}} &=& \frac{2v_{0}}{m_{+}} \ , \cr
&& \cr
\tau_{\scriptscriptstyle{2(k+1)}} - \tau_{\scriptscriptstyle{2k+1}}' \ = \ \tau_{\scriptscriptstyle{2k+1}}' - \tau_{\scriptscriptstyle{2k+1}} &=& \frac{1}{\xi v_{0}} \ . 
\end{eqnarray}
Observe that \ $x(\tau_{2k}) = n\pi + 2/\xi$ \ and \ $v_{0} = (dx/d\tau)(\tau_{2k})$ \ for all \ $k\in\mathbb{Z}$.

From (\ref{N15}) and (\ref{N14tiempos}) one finds that the bounded motion has period $P$ given by
\begin{eqnarray}
\label{N16}
P &=& \frac{2v_{0}}{m_{+}} + \frac{2}{\xi v_{0}} \ .
\end{eqnarray}
The absolute minimum value $P_{\mbox{\tiny min}}$ of $P$ occurs for \ $v_{0} = \sqrt{m_{+}/\xi} \simeq 1/\sqrt{2\xi}$ \ and is given by
\begin{eqnarray}
\label{Pmin}
P_{\mbox{\tiny min}} \ = \ \frac{4}{\sqrt{m_{+}\xi}} \ \simeq \ 4\sqrt{\frac{2}{\xi}} \ .
\end{eqnarray}
Here we used (\ref{91bis}) to conclude that \ $m_{+} \simeq 1/2$ \ for \ $\xi \gg 1$.

\begin{figure}[htbp]
  \centering
  \includegraphics[scale=0.75]{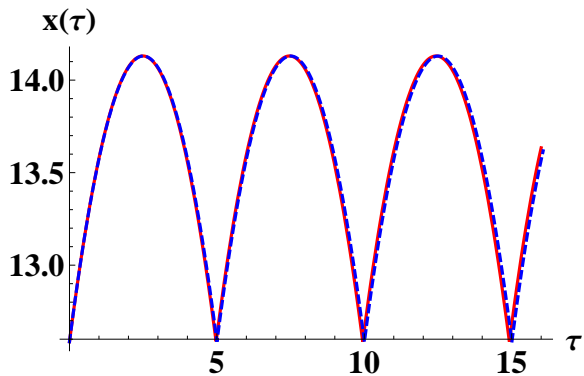}
  \caption{(Color online) The figure illustrates the solution $x(\tau)$ (red-solid line) of (\ref{N1RWA}) computed numerically for \ $\xi = 50$ \ and the initial conditions \ $x(0) = 4\pi + 1/\xi$, \ $x'(0) = 0$.  It also shows the approximate solution $x_{\mbox{\tiny RWA}}(\tau)$ (blue-dashed line) in (\ref{N15}) corresponding to the same energy \ $\mathcal{E} = E(4\pi + 1/\xi,0)$. The corresponding initial conditions for $x_{\mbox{\tiny RWA}}(\tau)$ are  \ $x(0) = 4\pi + 1/\xi$ \ and \ $x'(0) = x_{0}'(0)$ \ with $x_{0}'(0)$ in (\ref{CorrespondenciaCI}). }
\label{Figure6}
\end{figure}

Figure~\ref{Figure6} shows $x(\tau)$ (red-solid line) obtained by solving numerically (\ref{N1RWA}) for \ $\xi = 50$ \ and the initial conditions \ $x(0) = 4\pi + 1/\xi$ \ and \ $x'(0) = 0$. These initial conditions are such that the movable mirror starts from rest at a position where $\omega_{0}$ coincides with one of the cavity resonance frequencies. Also,\ $\mathcal{E} \equiv E(4\pi + 1/\xi,0)$ \ is the energy of the movable mirror. The figure also shows the approximate solution $x_{\mbox{\tiny RWA}}(\tau)$ (blue-dashed line) given in (\ref{N15}) and corresponding to the same energy $\mathcal{E}$. Notice that one has to adjust the initial velocity of the movable mirror in order to apply (\ref{N15}). According to the approximation in (\ref{91}) the corresponding initial conditions for $x_{\mbox{\tiny RWA}}(\tau)$ are \ $x(0) = 4\pi + 1/\xi$ \ and \ $x'(0) = x_{0}'(0)$ \ with
\begin{eqnarray}
\label{CorrespondenciaCI}
x_{0}'(0) &=& \sqrt{2\left[ \mathcal{E} - V_{\mbox{\tiny RWA}}\left( \frac{2}{\xi} \right) \right] } \ .
\end{eqnarray} 
Observe that the agreement between $x(\tau)$ and $x_{\mbox{\tiny RWA}}(\tau)$ is quite good. 
 
To end this subsection we determine sufficient conditions for (\ref{1condicionesEq}) to be valid. In the following we drop the assumption that \ $\xi \gg 1$ \ and we do not restrict the motion to be bounded.

From (\ref{N14}) we know that energy is conserved, that is, 
\begin{eqnarray}
\label{ValidezCampoi}
E[x(\tau),x'(\tau)]  &=& E_{0} \qquad (\tau\in\mathbb{R}) \ . 
\end{eqnarray}
From (\ref{N14}) and (\ref{ValidezCampoi}) one can solve for $x'(\tau)$ and then use the bounds in (\ref{CotasVRWA}) to conclude that
\begin{eqnarray}
\label{ValidezCampoii}
\vert x'(\tau) \vert &\leq& v_{\mbox{\tiny max}} \equiv \sqrt{2E_{0} + \pi }\ .
\end{eqnarray}
From the definition of $x(\tau)$ in (\ref{76}) one can relate derivatives of $x(\tau)$ with those of $q(t)$ as follows (recall that \ $\tau = \Delta t$):
\begin{eqnarray}
\label{CorrespondenciaX}
\frac{\dot{q}(t)}{c} \ = \ \frac{2}{\Omega}x'(\tau) \ , \qquad \frac{\ddot{q}(t)}{c\omega_{0}} \ = \ \frac{4}{\Omega^{2}}x''(\tau) \ .
\end{eqnarray}
Combining (\ref{ValidezCampoii}) with (\ref{CorrespondenciaX}) one obtains a sufficient condition for the first part of (\ref{1condicionesEq}) to be valid:
\begin{eqnarray}
\label{ValidezCampoiii}
2\sqrt{2E_{0} + \pi} \ll \Omega &\Rightarrow& \left\vert \frac{\dot{q}(t)}{c} \right\vert \ \ll \ 1 \qquad (t\in\mathbb{R}).
\end{eqnarray}
A sufficient condition for the second part of (\ref{1condicionesEq}) is given in (\ref{ValidezCampovi}). Note that (\ref{ValidezCampovi}) and (\ref{ValidezCampoiii}) essentially say that (\ref{1condicionesEq}) are satisfied if \ $\Omega \gg 1$.

To get an idea of the order of magnitude of the quantities involved in (\ref{ValidezCampovi}) and (\ref{ValidezCampoiii}) we take from Appendix III the typical  values \ $\xi = 6.4$ \ (which corresponds to a reflectivity of the movable mirror of $0.91$) and \ $\Omega \geq 10^{9}$. First observe from (\ref{MaximizadorfRWA}) that \ $8f_{\mbox{\tiny RWA}}(x_{2n}) \simeq 4\xi^{2} = 163.84 \ll 10^{18} \leq \Omega^{2}$. From (\ref{ValidezCampovi}) one then obtains that \ $\vert \ddot{q}(t)/(c\omega_{0}) \vert \ll 1$ \ for all \ $t\in\mathbb{R}$. Notice that this condition is valid for both bounded and unbounded motion. Now observe that \ $2\sqrt{2E_{0} + \pi} \leq 2\sqrt{\pi} \ll 10^{9} \leq \Omega$ \ for bounded motion. From (\ref{ValidezCampoiii}) it follows that \ $\vert \dot{q}(t)/c \vert \ll 1$ \ for all \ $t\in\mathbb{R}$ \ if only bounded motion is considered. Therefore, (\ref{1condicionesEq}) are satisfied. 

If one considers only bounded motion (that is, \ $E_{0} \leq 0$) and that the transparency of the movable mirror is very small (that is, \ $\xi \gg 1$), then it follows from (\ref{ValidezCampovi}) and (\ref{ValidezCampoiii}) that
\begin{eqnarray}
\label{ValidezCampovii}
4\xi^{2} \ll \Omega^{2} &\Rightarrow& \left\vert \frac{\ddot{q}(t)}{c \omega_{0}} \right\vert , \  \left\vert \frac{\dot{q}(t)}{c} \right\vert \ \ll \ 1 \ \ \ (t\in\mathbb{R}).
\end{eqnarray}
Here we have used \ $f_{\mbox{\tiny RWA}}(x_{2n}) \simeq \xi^{2}/2$, see (\ref{MaximizadorfRWA}). To illustrate the order of magnitude of the quantities in (\ref{ValidezCampovii}) we take \ $\xi = 50$ \ (which corresponds to a reflectivity of the movable mirror of $0.9984$) and  \ $\Omega \geq 10^{9}$. It follows that \ $4\xi^{2} = 10^{4} \ll 10^{18} \leq \Omega^{2}$. It follows from (\ref{ValidezCampovii}) that the conditions in (\ref{1condicionesEq}) are satisfied.

Finally, we rewrite (\ref{CondicionIntensidadDebil}) for the case of  bounded motion, that is, \ $E_{0} \leq 0$. The condition for the RWA regime is given by (\ref{CondicionIntensidadDebil}) but with $P$ the period of the bounded motion. In the case \ $\xi \gg 1$ \ (that is, the transparency of the movable mirror is very small) one can use (\ref{Pmin}) to obtain a general sufficient condition for the RWA to be valid:
\begin{eqnarray}
\label{CondicionIntensidadDebilXigg1}
\Omega &\gg& \frac{\pi}{2}\sqrt{m_{+}\xi} \ \simeq \ \frac{\pi}{2}\sqrt{\frac{\xi}{2}}  \ . 
\end{eqnarray}
Comparing (\ref{CondicionIntensidadDebilXigg1}) with (\ref{ValidezCampovii}) one confirms that the low intensity limit (or RWA regime) is the natural setting for the consideration of the model established in the first section in the case where the transparency of the movable mirror is small (that is, \ $\xi \gg 1$).

\subsection{The intermediate case}

\begin{figure}[htbp]
  \centering
  \includegraphics[scale=0.75]{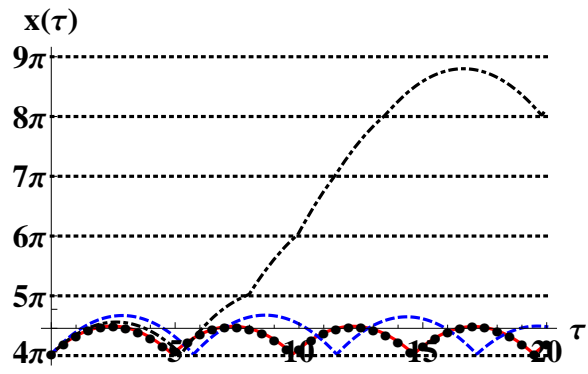}\\
  \caption{(Color online) 
The figure illustrates $x(\tau)$ computed numerically from (\ref{80}) with \ $\Omega = 100$ (black-dotdashed line), $250$ (blue-dashed line), and $500$ (red-solid line) for \ $\xi = 10$ \ and the initial conditions \ $x(0) = 4\pi + 1/\xi$ \ and \ $x'(0) = 0$. The horizontal black-dotted lines indicate the beginning and the ends of the potential wells of $V_{\mbox{\tiny RWA}}(x)$. The figure also illustrates the solution in the RWA (black-dotted line) calculated numerically from (\ref{N1RWA}).}
\label{Figure7}
\end{figure}

In this subsection we consider equation (\ref{80}) without making any assumptions on $\Omega$. In this case the mirror does not have conservative dynamics because the force affecting it depends explicitly on time. As a consequence, the dynamics are much more complex in this case. If the movable mirror is initially confined to one of the wells of $V_{\mbox{\tiny RWA}}(x)$, at future times it may jump out of the well and then be confined for some time in another well. Also, oscillations in one well similar to those obtained with the RWA in the previous subsection are possible for large enough $\Omega$. Figure~\ref{Figure7} illustrates these types of behaviors for several values of $\Omega$ and compares them with the solution in the RWA (black-dotted line) calculated numerically from (\ref{N1RWA}).  Notice how $x(\tau)$ tends to the solution in the RWA for increasing $\Omega$. 

We now obtain sufficient conditions for (\ref{1condicionesEq}) to be valid. A sufficient condition for the acceleration of the movable mirror is given in (\ref{ValidezCampovi}). We illustrate it using the parameters of Fig.~\ref{Figure7}. Since \ $\xi = 10$ \ and \ $\Omega \geq 100$, one has from (\ref{MaximizadorfRWA}) that \ $8f_{\mbox{\tiny RWA}}(x_{2n}) \simeq 400 \ll 10^{4} \leq \Omega^{2}$. From (\ref{ValidezCampovi}) it follows that the second condition in (\ref{1condicionesEq}) is satisfied for the parameters of Fig.~\ref{Figure7}.

We now consider the velocity of the movable mirror. If \ $\vert x'(\tau)\vert \leq v_{\mbox{\tiny max}}$ \ for all $\tau$, it follows from (\ref{CorrespondenciaX}) that   
\begin{eqnarray}
\label{ValidezCampoNorwa3}
2v_{\mbox{\tiny max}} \ll \Omega &\Rightarrow& \left\vert \frac{\dot{q}(t)}{c} \right\vert \ll 1 \qquad (t\in\mathbb{R}),
\end{eqnarray}
The problem here is that there is no direct way to bound the velocity as in the case of the RWA regime because energy is no longer conserved. We now illustrate (\ref{ValidezCampoNorwa3}) using the parameters of Fig.~\ref{Figure7}, that is, \ $\xi = 10$ \ and \ $\Omega \geq 100$. All the numerical solutions in Fig.~\ref{Figure7} have \ $\vert x'(\tau) \vert \leq 2.5$ \ for the $\tau$ interval shown, so that \ $2v_{\mbox{\tiny max}} \leq 5 \ll 100 \leq \Omega$. Hence, it follows from (\ref{ValidezCampoNorwa3}) that the first condition in (\ref{1condicionesEq}) is satisfied for the parameters of Fig.~\ref{Figure7}.

To end this subsection we note that (\ref{91}) can be used to solve (\ref{80}) approximately if \ $\xi \gg 1$. We do not pursue this direction here.

\section{INTRODUCTION OF FRICTION}

In this section we assume that the movable mirror is subject to friction linear in the velocity. For example, one could think that the mirror is attached to a set of wheels that subjects it to friction and allows it to move due to the radiation pressure exerted by the field. In this case the equation of motion for the mirror is given by
\begin{eqnarray}
\label{F1}
&& \frac{d^{2}x}{d\tau^{2}}(\tau) + \Gamma \frac{dx}{d\tau}(\tau) \cr
&=& \left\{ 1 + \mbox{cos}\left[ \Omega\tau + 2x(\tau) -2\delta_{k_{\scriptscriptstyle{N}}^{\scriptscriptstyle{0}}}\left( \frac{x(\tau)}{k_{\scriptscriptstyle{N}}^{\scriptscriptstyle{0}}} \right) \right] \right\} \times \cr
&& \qquad \times f_{\mbox{\tiny RWA}}\left[ x(\tau) \right] \ . 
\end{eqnarray}  
We note that (\ref{F1}) is obtained by adding on the right of (\ref{2}) the friction force \ $-\gamma \dot{q}(t)$ \ and then using (\ref{76}) to express the equation in terms of non-dimensional quantities. Here $\gamma$ has units of Ns/m$^{3}$ and \ $\Gamma = \gamma/(\Delta M_{0})$ \ is a non-dimensional quantity. 

As before, we shall consider the RWA regime and the intermediate case separately.

\subsection{The limit of low field intensity}

The limit of low field intensity is still defined by condition (\ref{CondicionIntensidadDebil}) and it allows one to perform the RWA in (\ref{F1}):
\begin{eqnarray}
\label{F45}
\frac{d^{2}x}{d\tau^{2}}(\tau) + \Gamma \frac{dx}{d\tau}(\tau) &=& f_{\mbox{\tiny RWA}}\left[ x(\tau) \right] \ . \ \
\end{eqnarray}

We now perform a linear stability analysis of (\ref{F45}). The aforementioned equation can be written as a first order non-linear system as follows:
\begin{eqnarray}
\label{F46sistema}
\frac{d}{d\tau} 
\left(
\begin{matrix}
x(\tau) \cr
y(\tau)
\end{matrix}
\right)
&=& \mathbf{f}\left[ x(\tau),y(\tau) \right] \ , \cr
&\equiv&
\left(
\begin{matrix}
y(\tau) \cr
-\Gamma y(\tau) + f_{\mbox{\tiny RWA}}\left[ x(\tau) \right] 
\end{matrix}
\right) \ ,
\end{eqnarray}
with \ $y(\tau) = x'(\tau)$. Using (\ref{ii4}) it follows that the fixed points of (\ref{F46sistema}) are determined by the maximizers and minimizers of $V_{\mbox{\tiny RWA}}(x)$. Explicitly, the fixed points of (\ref{F46sistema}) are given by \ $(x_{n}^{*},y_{n}^{*}) = (n\pi, 0)$ \ and \ $(x_{n}^{**},y_{n}^{**}) \simeq (n\pi + 2/\xi, 0)$ \ with \ $n\in\mathbb{Z}^{+}$, see (\ref{MaximizadorVRWA}) and (\ref{MinimizadorVRWA}) .

The linearization of (\ref{F46sistema}) at the fixed points reveals that \ $(x_{n}^{*},y_{n}^{*}) = (n\pi, 0)$ \ are saddle points (see Theorem 3 in Section 2.10 of \cite{Strogatz}) and that \ $(x_{n}^{**},y_{n}^{**}) \simeq (n\pi + 2/\xi, 0)$ \ are attractors (see Theorem 4 in Section 2.10 of \cite{Strogatz}). 

Recall that a fixed point is an attractor if there is a neighborhood $V$ of the fixed point such that any phase space trajectory $(x(\tau),y(\tau))$ that starts in $V$ tends to the fixed point as $\tau \rightarrow + \infty$. 

In particular, the attractors are stable spirals if
\begin{eqnarray}
\label{F46espiral}
\frac{\Gamma}{2} &<& \sqrt{\vert f_{\mbox{\tiny RWA}}'(x_{n}^{**}) \vert} \ \simeq \ \sqrt{\xi} \ .
\end{eqnarray}
On the other hand, the attractors are stable nodes if 
\begin{eqnarray}
\label{F46nodo}
\frac{\Gamma}{2} &\geq& \sqrt{\vert f_{\mbox{\tiny RWA}}'(x_{n}^{**}) \vert} \ \simeq \ \sqrt{\xi} \ .
\end{eqnarray}
(See Theorem 4 of Section 2.10 of \cite{Strogatz}). We note that in (\ref{F46espiral}) and (\ref{F46nodo}) the value on the extreme right is obtained by neglecting terms of order $1/\xi^{2}$ and smaller with respect to $1$.

Recall that a fixed point is called a stable spiral or focus if there is a deleted neighborhood $V$ of the fixed point such that every phase space trajectory $(x(\tau),y(\tau))$ in $V$ \textit{spirals toward the fixed point} as $\tau \rightarrow +\infty$. On the other hand, a fixed point is called a stable node if there is deleted neighborhood $V$ of the fixed point such that every phase space trajectory $(x(\tau),y(\tau))$ in $V$ approaches the fixed point along a well-defined tangent line as $\tau \rightarrow + \infty$.

\begin{figure}[htbp]
  \centering
  \subfloat[]{\label{Figure8a}\includegraphics[scale=0.75]{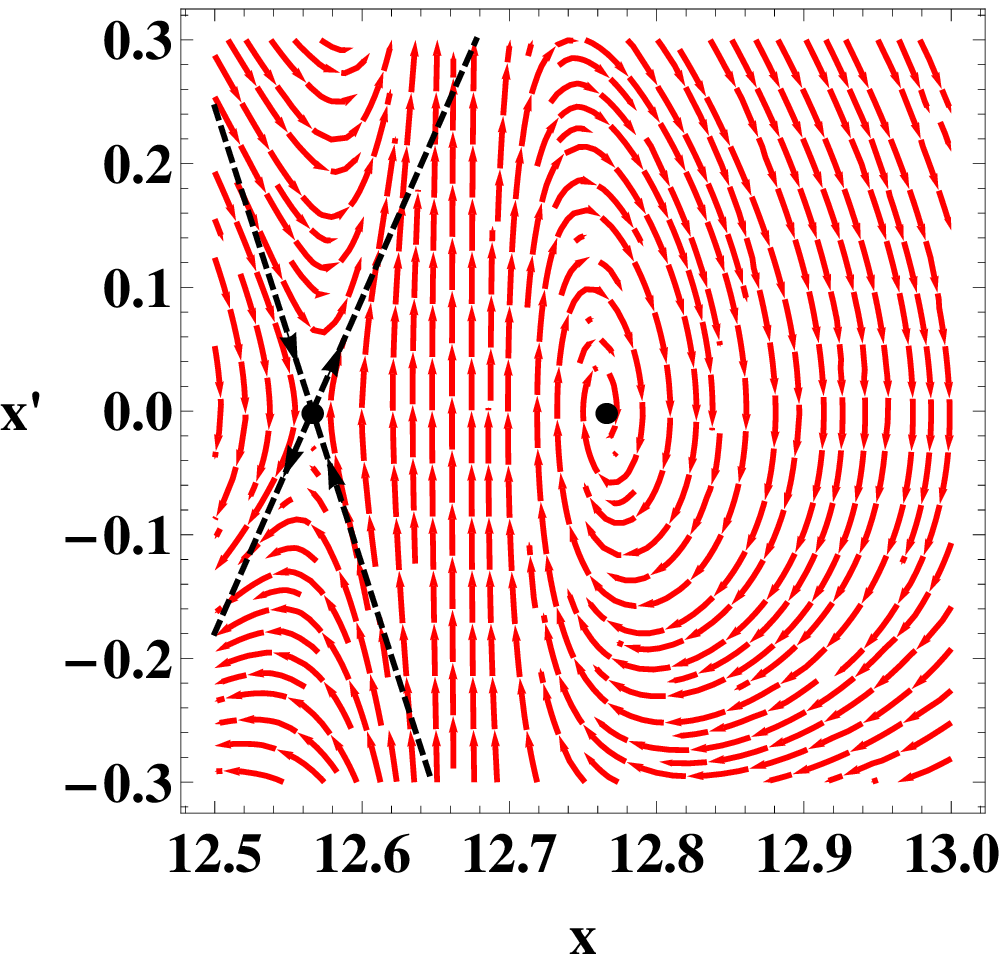}} \hspace{0.3cm}
  \subfloat[]{\label{Figure8b}\includegraphics[scale=0.75]{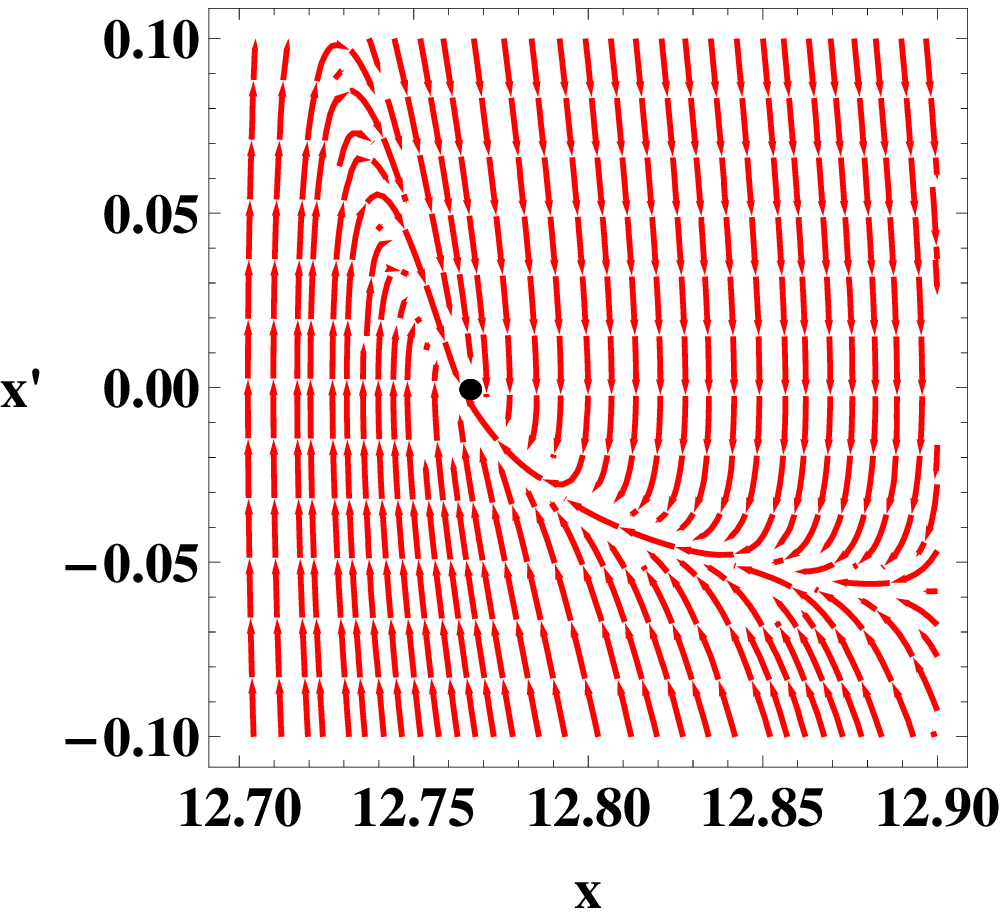}} \\
  \caption{(Color online) Figure~\ref{Figure8a} illustrates the phase portrait of the movable mirror in the case \ $\xi = 10$ \ and \ $\Gamma = 1$. The black dot on the left indicates the saddle point \ $(x_{n}^{*},y_{n}^{*}) = (4\pi ,0)$, while the black dot on the right indicates the stable spiral \ $(x_{n}^{**},y_{n}^{**}) \simeq (4\pi + 2/\xi, 0)$. Also, the black-dashed line with negative (positive) slope indicates the stable (unstable) manifold of the saddle point. Figure~\ref{Figure8b} illustrates the phase portrait of the movable mirror in the case \ $\xi = 10$ \ and \ $\Gamma = 7$. The black dot indicates the stable node \ $(x_{n}^{**},y_{n}^{**}) \simeq (4\pi + 2/\xi, 0)$.}
\label{Figure8}
\end{figure}

Figure~\ref{Figure8} illustrates these different types of behaviors for \ $\xi = 10$. Notice that the movable mirror exhibits a behavior similar to an under-damped harmonic oscillator in the case of the stable spiral and to an over-damped harmonic oscillator in the case of a stable node.

Figure~\ref{Figure8a} also illustrates the stable and unstable manifolds of a saddle point calculated using the linearization of (\ref{F46sistema}). They are given by
\begin{eqnarray}
\label{StableUnstableManifoldF}
x' &=& \lambda_{\pm}(x - n\pi) \ ,
\end{eqnarray}
with \ $n \in\mathbb{Z}^{+}$ \ and $\lambda_{\pm}$ the eigenvalues of the coefficient matrix of the linearized system at the saddle point 
\begin{equation}
\label{lambdapm}
\lambda_{\pm} \ = \ -\frac{\Gamma}{2} \pm \sqrt{ \left(\frac{\Gamma}{2}\right)^{2} + \xi} \ . \ \
\end{equation}
The stable (unstable) manifold corresponds to the minus (plus) sign. 

Recall that the stable manifold of a saddle point is the set of initial conditions such that $(x(\tau),y(\tau))$ tends to the saddle point as $\tau\rightarrow + \infty$, while the unstable manifold of a saddle point is the set of initial conditions such that $(x(\tau),y(\tau))$ tends to the saddle point as $\tau\rightarrow - \infty$. 

Since \ $\nabla \cdot \mathbf{f}(x,y) = -\Gamma < 0$ \ for all \ $(x,y)\in\mathbb{R}^{+}\times\mathbb{R}$ \ with $\mathbf{f}(x,y)$ in (\ref{F46sistema}), it follows from Bendixon's criterion (see Theorem 1 in Section 3.9 of \cite{Strogatz}) that there are no closed orbits and no cycle-graphs in phase space. 

Recall that a cycle-graph is also called a compound separatrix cycle and that it is a closed curve in phase space that consists of \ $N \geq 1$ \ vertices and at least $N$ edges. The vertices are fixed points and the edges are trajectories that tend to vertices as \ $\tau \rightarrow \pm \infty$. The whole curve is coherently oriented by increasing time. For a precise statement see Definition 1 in Section 3.7 of \cite{Strogatz}. 

The non-existence of closed orbits and cycle-graphs in phase space could have also been anticipated by noting that the energy of the mirror defined in (\ref{N14}) is always decreasing:
\begin{eqnarray}
\label{F46energia}
E\left[ x(\tau), x'(\tau) \right] &=& E\left[ x(0), x'(0) \right] - \Gamma\int_{0}^{\tau}d\tau' \left[ x'(\tau') \right]^{2} \ . \cr
&&
\end{eqnarray}

For \ $\tau \gg 1$ \ the movable mirror always tends to the position $x_{n}^{**}$ of a minimizer of $V_{\mbox{\tiny RWA}}(x)$. The only exceptions are the trajectories associated with the stable manifolds of the saddle points in which case the movable mirror tends to a maximizer $x_{n}^{*}$ of $V_{\mbox{\tiny RWA}}(x)$. The behavior stated in the previous lines is a consequence of the Poincar\'{e}-Bendixon theory in $\mathbb{R}^{2}$ (see Section 3.7 of \cite{Strogatz}) and the facts that there are no closed orbits and no cycle-graphs in phase space and that the energy of the mirror is always decreasing (so that trajectories in phase space are bounded for $\tau \geq 0$). 

Taking \ $\xi \gtrsim 5$ \ and using (\ref{ExtraMaximizadoresQ2n}), (\ref{ExtraLorentzianaL}), and (\ref{MaximizadorfRWA}) one finds that the frequency $\omega_{0}$ of the field approximately coincides with one of the cavity resonance frequencies if only if the non-dimensional position \ $x = k_{\scriptscriptstyle{N}}^{\scriptscriptstyle{0}}q$ \ of the movable mirror satisfies \ $\vert x -x_{2n} \vert \leq 1/\xi^{2}$ \ with \ $x_{2n} = k_{\scriptscriptstyle{N}}^{0}q_{2n} \simeq n\pi + 1/\xi$. For \ $\xi \gtrsim 5$ \ one has \ $\vert x -x_{2n} \vert \simeq 1/\xi \gg 1/\xi^{2}$ \ if \ $x = x_{n}^{*} = n\pi$ \ or \ $x= x_{n}^{**} \simeq n\pi + 2/\xi$, see (\ref{MaximizadorVRWA}) and (\ref{MinimizadorVRWA}). Therefore, the movable mirror ends up in a position that is very different from one in which $\omega_{0}$ coincides with one of the cavity resonance frequencies if \ $\xi \gtrsim 5$.

We now solve analytically (\ref{F45}) to good approximation in the case \ $\xi \gg 1$ \ and \ $E[x(0),x'(0)] <0$. Using the approximations given in (\ref{91}) and (\ref{93}) it follows that the solution $x_{\mbox{\tiny RWA}}(\tau)$ of the aforementioned equation is
\begin{eqnarray}
\label{F47}
x_{\mbox{\tiny RWA}}(\tau) &=& 
c_{2}(\tau_{0})e^{-\Gamma (\tau - \tau_{0})} - \frac{m_{+}}{\Gamma}(\tau -\tau_{0}) \cr 
&& \cr
&& + c_{1}(\tau_{0})  + \frac{m_{+}}{\Gamma^{2}}\left[ 1 -e^{-\Gamma (\tau -\tau_{0})} \right] \ , \ \ \ \ 
\end{eqnarray}
with
\begin{eqnarray}
\label{F47const}
c_{1}(\tau_{0}) &=& x_{\mbox{\tiny RWA}}(\tau_{0}) + \frac{1}{\Gamma}x_{\mbox{\tiny RWA}}'(\tau_{0}) \ , \cr
c_{2}(\tau_{0}) &=& -\frac{1}{\Gamma}x_{\mbox{\tiny RWA}}'(\tau_{0}) \ ,
\end{eqnarray} 
if \ $(n\pi + 2/\xi) \leq x_{\mbox{\tiny RWA}}(\tau) \leq (n+1)\pi$ \ for \ $\tau_{0} \leq \tau \leq \tau_{1}$ \ and some \ $n\in\mathbb{Z}^{+}$. Also,
\begin{eqnarray}
\label{NuevaFriccion}
x_{\mbox{\tiny RWA}}(\tau) &=& x_{\mbox{\tiny RWA}}(\tau_{1}) + \frac{x_{\mbox{\tiny RWA}}'(\tau_{1})}{\Gamma}\left[ 1- e^{-\Gamma(\tau-\tau_{1})} \right] \ , \ \ \
\end{eqnarray}
if \ $(n\pi + 1/\xi) \leq x_{\mbox{\tiny RWA}}(\tau) \leq (n\pi + 2/\xi)$ \ for \ $\tau_{1} \leq \tau \leq \tau_{2}$. In order to get the complete trajectory, the solutions in (\ref{F47}) and (\ref{NuevaFriccion}) have to be pasted together in order to guarantee the continuity of $x_{\mbox{\tiny RWA}}(\tau)$ and $x_{\mbox{\tiny RWA}}'(\tau)$. Moreover, it has to be taken into account that $x_{\mbox{\tiny RWA}}(\tau)$ bounces elastically from a potential wall if it reaches \ $(n\pi + 1/\xi)$ \ (see the approximation in (\ref{91})). We omit the details.

\begin{figure}[htbp]
  \centering
  \subfloat[]{\label{Figure9a}\includegraphics[scale=0.75]{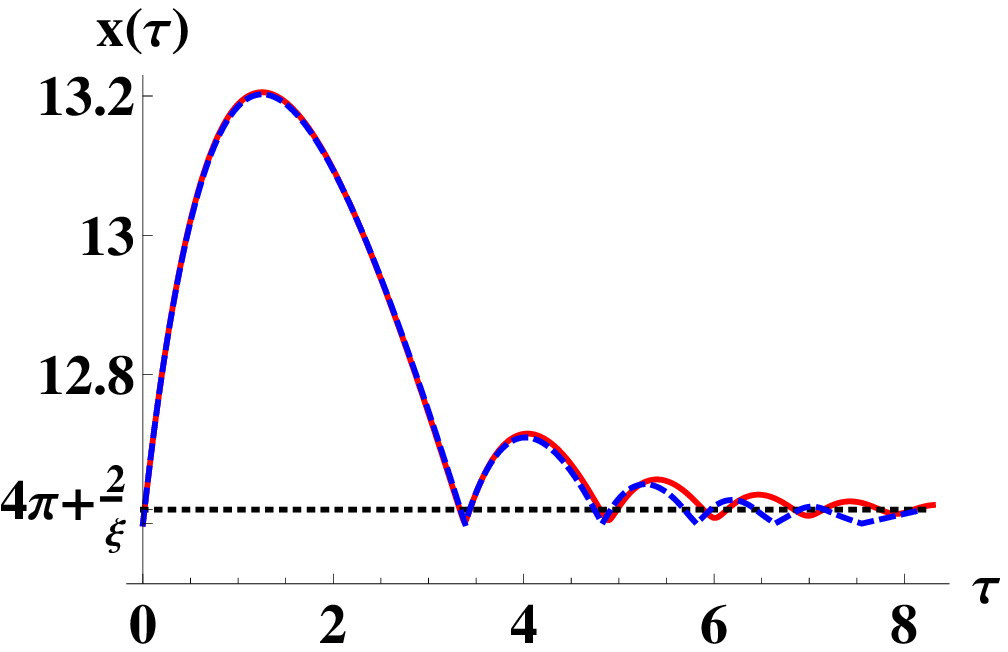}} \hspace{0.3cm}
  \subfloat[]{\label{Figure9b}\includegraphics[scale=0.75]{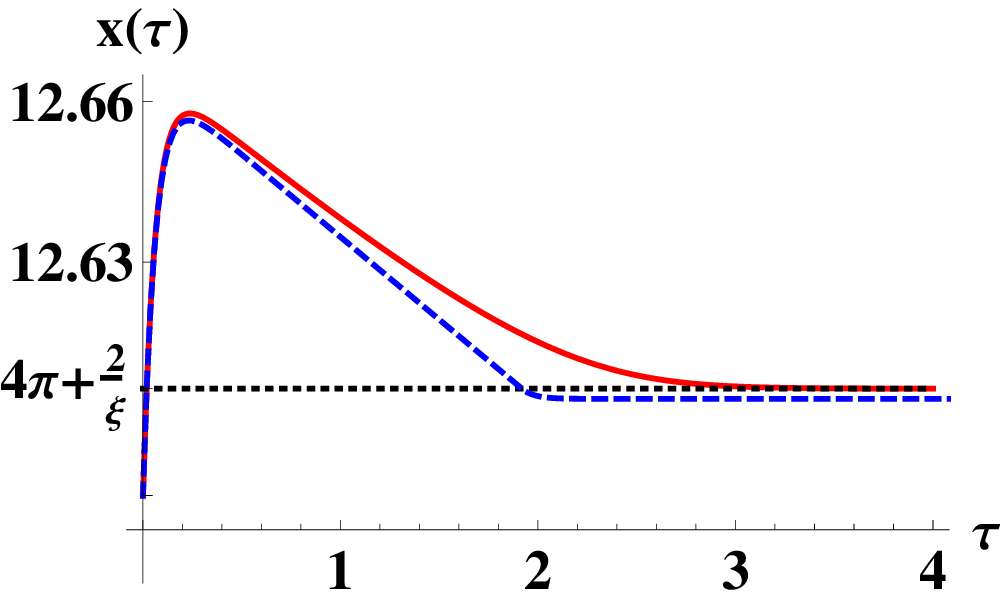}} \\
  \caption{(Color online) The figures compare the numerical solution (red-solid line) of (\ref{F45}) and the approximate $x_{\mbox{\tiny RWA}}(\tau)$ in (\ref{F47}) and (\ref{NuevaFriccion}) (blue-dashed line) for $\xi = 50$ and the initial conditions \ $x(0) = (4\pi + 1/\xi)$ \ and \ $x'(0) = 0$. Fig.~\ref{Figure9a} has \ $\Gamma = 1$, while Fig.~\ref{Figure9b} has $\Gamma = 16$.}
\label{Figure9}
\end{figure}

Figure~\ref{Figure9} compares the approximate $x_{\mbox{\tiny RWA}}(\tau)$ with $x(\tau)$ computed numerically from (\ref{F45}) for \ $\xi = 50$ \ and two values of $\Gamma$. The initial conditions are \ $x(0) = (4\pi +1/\xi)$ \ and \ $x'(0) = 0$, so that the movable mirror starts from rest at a position where $\omega_{0}$ coincides with one of the cavity resonance frequencies. As in the previous section, in order to use (\ref{F47}) and (\ref{NuevaFriccion}) one has to adjust the initial velocity $x'(0)$ to be $x_{0}'(0)$ in (\ref{CorrespondenciaCI}) with \ $\mathcal{E} = E(4\pi +1/\xi,0)$ \ the initial energy. Notice that the agreement between the numerical solution and $x_{\mbox{\tiny RWA}}(\tau)$ is quite good for not very large times in Fig.~\ref{Figure9a} which corresponds to the case where \ $\Gamma = 1$ \ and \ $(x_{n}^{**},y_{n}^{**}) \simeq (4\pi + 2/\xi , 0)$ \ is a stable spiral. The agreement is not so good in Fig.~\ref{Figure9b} which corresponds to the case where \ $\Gamma = 16$ \ and \ $(x_{n}^{**},y_{n}^{**}) \simeq (4\pi + 2/\xi , 0)$ \ is a stable node. The differences between the analytical and the numerical solutions are due to two facts. First, the approximation in (\ref{91}) does not take into account the curvature of the potential $V_{\mbox{\tiny RWA}}(x)$. Second, for large enough times the movable mirror spends more time near the minimizer $x_{n}^{**}$ of $V_{\mbox{\tiny RWA}}(x)$ where the curvature is important. Nevertheless, $x_{\mbox{\tiny RWA}}(\tau)$ allows one to understand the behavior of the mirror. From (\ref{F47}) it follows that the movable mirror approximately behaves like a particle in free fall subject to friction (linear in the velocity) when it is in the region between the minimizer \ $x_{n}^{**} \simeq (n\pi + 2/\xi)$ \ and the maximizer \ $x_{n+1}^{*} = (n+1)\pi$ \ of $V_{\mbox{\tiny RWA}}(x)$. On the other hand, from (\ref{NuevaFriccion}) we see that it behaves approximately like a particle subject only to friction in the region between the maximizer \ $x_{2n} \simeq (n\pi +1/\xi)$ \ of $f_{\mbox{\tiny RWA}}(x)$ and $x_{n}^{**}$. Moreover, the mirror bounces elastically from an impenetrable potential wall if it reaches $x_{2n}$, which corresponds to the maximizer of $f_{\mbox{\tiny RWA}}(x)$ (see (\ref{MaximizadorfRWA}) and (\ref{91})).

To end this subsection we give sufficient conditions for (\ref{1condicionesEq}) to be valid. In the following the discussion is general and we do not assume that \ $\xi \gg 1$ \ or that \ $E[x(0),x'(0)] < 0$.

From (\ref{N14}) and (\ref{F46energia}) it follows that (\ref{ValidezCampoii}) still holds with \ $E_{0} = E[x(0),x'(0)]$ \ and \ $\tau \geq 0$. Hence, (\ref{ValidezCampoiii}) is still valid with \ $E_{0} = E[x(0),x'(0)]$ \ and \ $t \geq 0$.

Using (\ref{ValidezCampoii}) with \ $E_{0} = E[x(0),x'(0)]$ \ and \ $\tau \geq 0$ \ in combination with (\ref{CorrespondenciaX}) and (\ref{F45}) it follows that 
\begin{eqnarray}
\label{ValidezFriccionv}
\left\vert \frac{\ddot{q}(t)}{c\omega_{0}} \right\vert 
&\leq&  \frac{f_{\mbox{\tiny RWA}}(x_{2n}) + \Gamma \sqrt{2E_{0}+\pi}}{( \Omega /2)^{2}} \ \ \ \ (t \geq 0). \ \ \ \ 
\end{eqnarray}
Here we have used that $x_{2n}$ is a maximizer of $f_{\mbox{\tiny RWA}}(x)$, see (\ref{MaximizadorfRWA}). Hence,
\begin{eqnarray}
\label{ValidezFriccionvi}
&&  4f_{\mbox{\tiny RWA}}(x_{2n}) + 4\Gamma \sqrt{2E_{0}+\pi} \ \ll \Omega^{2} \cr
&& \cr
&\Rightarrow& \left\vert \frac{\ddot{q}(t)}{c\omega_{0}} \right\vert \ll 1 \ \ \ (t \geq 0). \ \ \ \
\end{eqnarray}
Notice that (\ref{ValidezFriccionvi}) is similar to (\ref{ValidezCampovi}), since it only adds the term $4\Gamma \sqrt{2E_{0}+\pi}$. 

As in the case of dynamics due only to radiation pressure, conditions (\ref{ValidezCampoiii}) and (\ref{ValidezFriccionvi}) essentially say that the model applies if $\Omega$ is sufficiently large.  Using the values \ $\Gamma/ \Omega \sim 10^{-14}$ \ and \ $\Omega \gtrsim 10^{9}$ \ from Appendix III, it follows that the conditions for the validity of the model presented in Sec. II with the addition of friction (linear in the velocity) basically reduce to those of the case where there is no friction discussed in the previous section if $E_{0}$ is not too large.

If \ $\xi \gg 1$ \ (that is, the transparency of the movable mirror is very small) and \ $E_{0} \leq 0$ \ (that is, the movable mirror is initially bounded to one of the potential wells), then conditions (\ref{ValidezCampoiii}) and (\ref{ValidezFriccionvi}) reduce to the following:
\begin{eqnarray}
\label{ValidezFriccionvii}
&& 4\Gamma\sqrt{\pi} + 2\xi^{2} \ \ll \ \Omega^{2} \cr
&& \cr
&\Rightarrow& \left\vert \frac{\dot{q}(t)}{c} \right\vert , \ \left\vert \frac{\ddot{q}(t)}{c\omega_{0}} \right\vert \ll 1 \ \ \ (t \geq 0).
\end{eqnarray}
Here we used (\ref{MaximizadorfRWA}) to evaluate $f_{\mbox{\tiny RWA}}(x_{2n})$.

Finally, we assume that $\xi \gg 1$ (that is, the transparency of the movable mirror is very small) and \ $E_{0} < 0$ \ (that is, the movable mirror starts in one of the potential wells at time $t=0$). Since the effect of friction is to slow down the movable mirror to an eventual stop, it follows that the time in which $x(\tau)$ changes appreciably can be estimated by the time in which $x(\tau)$ changes appreciably without friction. Hence, one can take (\ref{CondicionIntensidadDebilXigg1}) to be a sufficient condition for the RWA to be valid.
Again, comparing (\ref{CondicionIntensidadDebilXigg1}) and (\ref{ValidezFriccionvii}) it follows that the low intensity limit (or RWA regime) is a natural setting for the model presented in the first section.


\subsection{The intermediate case}

\begin{figure}[htbp]
  \centering
  \includegraphics[scale=0.75]{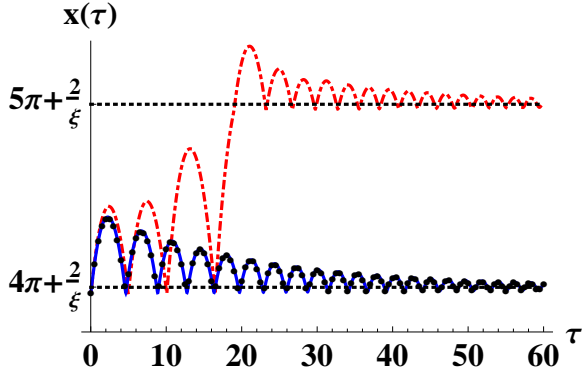}\\
  \caption{(Color online) The figure shows $x(\tau)$ computed numerically from (\ref{F1}) for \ $\xi=10$, \ $\Gamma = 0.07$, \ $\Omega = 100$ (blue-solid line), $\Omega = 500$ (red-dotdashed line), and the initial conditions \ $x(0) = 4\pi + 1/\xi$ \ and \ $x'(0) = 0$. It also shows $x_{\mbox{\tiny RWA}}(\tau)$ (black-dotted line) computed numerically from (\ref{F45}) and the horizontal black-dotted lines $(4\pi +2/\xi)$ and $(5\pi +2/\xi)$ showing approximately the position of minimizers $x_{n}^{**}$ of $V_{\mbox{\tiny RWA}}(x)$.}
\label{Figure10}
\end{figure}

\begin{figure}[htbp]
  \centering
  \subfloat[]{\label{Figure11a}\includegraphics[scale=0.75]{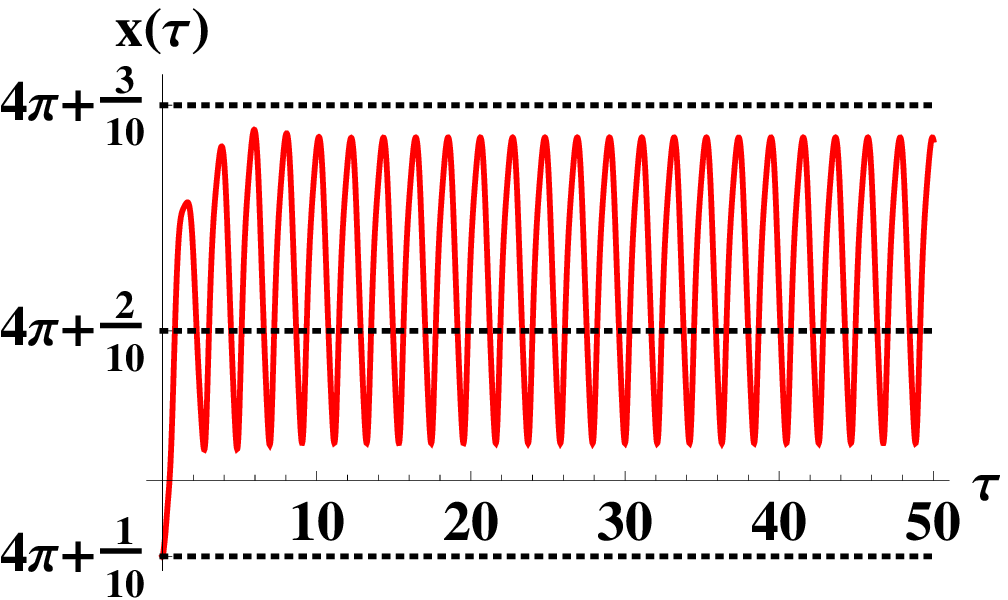}} \hspace{0.3cm}
  \subfloat[]{\label{Figure11b}\includegraphics[scale=0.75]{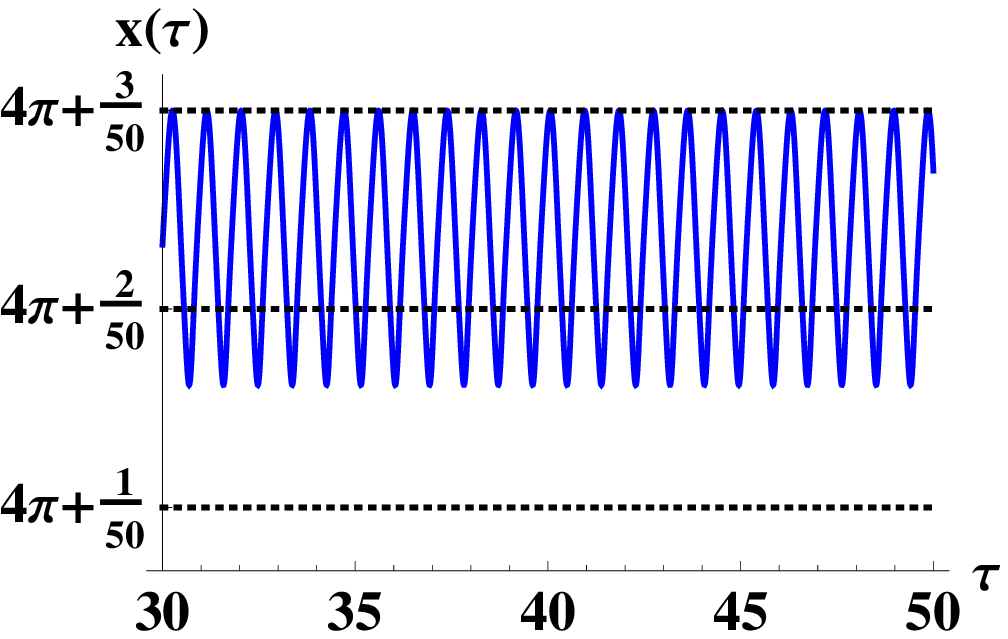}} \hspace{0.3cm} \\
  \caption{(Color online) The figures show $x(\tau)$ computed numerically from (\ref{F1}) for \ $\Gamma = 1$ \ and the initial conditions \ $x(0) = 4\pi + 1/\xi$ \ and \ $x'(0) = 0$. Figure \ref{Figure11a} has \ $\xi = 10$ \ and \ $\Omega = 6$, while figure \ref{Figure11b} has \ $\xi = 50$ \ and \ $\Omega = 14.1$. The figures also show horizontal black-dotted lines indicating the the position of the point $(n\pi+3/\xi)$ and the approximate positions of the maximizer \ $x_{2n} \simeq n\pi + 1/\xi$ \ of $f_{\mbox{\tiny RWA}}(x)$ and of the minimizer \ $x_{n}^{**} \simeq n\pi + 2/\xi$ \ of $V_{\mbox{\tiny RWA}}(x)$.}
\label{Figure11}
\end{figure}

In this subsection we go back to (\ref{F1}), where the movable mirror does not have conservative dynamics because the force affecting it depends explicitly on time. As before, the dynamics of the mirror are much more complex in this case. The numerical results show that depending on the values of $\Omega$, $\Gamma$, and $\xi$, the mirror can simply tend to a minimizer of $V_{\mbox{\tiny RWA}}(x)$ in a similar way as before or, if the friction is small enough, it can jump between potential wells before finally settling in one of the minimizers, see Fig.~\ref{Figure10}. Even periodic trajectories around a minimizer of $V_{\mbox{\tiny RWA}}(x)$ seem to exist, see Fig.~\ref{Figure11}. These periodic trajectories were obtained by taking \ $\Omega/2 \simeq \sqrt{\xi}(1+16/(3\xi^{2}))^{-1} \simeq \vert f_{\mbox{\tiny RWA}}'(x_{n}^{**}) \vert^{1/2}$ \ (compare this condition with (\ref{F46espiral}) and (\ref{F46nodo})). It is an interesting open question to characterize and actually prove the existence of these periodic trajectories. In figures \ref{Figure10} and \ref{Figure11} $x(\tau)$ was computed numerically from (\ref{F1}) for several values of $\xi$, $\Gamma$, and $\Omega$ and for initial conditions such that the mirror starts from rest at a position where $\omega_{0}$ coincides with one of the cavity resonance frequencies. Figure~\ref{Figure10} also shows $x_{\mbox{\tiny RWA}}(\tau)$ calculated numerically from (\ref{F45}). Notice that $x(\tau)$ with \ $\Omega = 500$ \ is practically the same as $x_{\mbox{\tiny RWA}}(\tau)$. Also, we have obtained numerical evidence that the dynamics of the mirror are very sensitive to the values of the parameters, since small changes can lead to very different dynamics. 

Using (\ref{CorrespondenciaX}) and (\ref{F1}) it can be shown that sufficient conditions for (\ref{1condicionesEq}) with \ $t\geq 0$ \ are the following:
\begin{eqnarray}
\label{ValidezFriccionSinRWA1}
2v_{\mbox{\tiny max}} \ \ll \ \Omega \ , \qquad 4\Gamma v_{\mbox{\tiny max}} + 8f_{\mbox{\tiny RWA}}(x_{2n})\ \ll \ \Omega^{2} \ ,
\end{eqnarray}
with $v_{\mbox{\tiny max}}$ an upper bound of $ \vert x'(\tau) \vert$ for \ $\tau \geq 0$ \ and $x_{2n}$ a maximizer of $f_{\mbox{\tiny RWA}}(x)$. If \ $\xi \gg 1$, then these reduce to the following:
\begin{eqnarray}
\label{ValidezFriccionSinRWA2}
2v_{\mbox{\tiny max}} \ \ll \ \Omega \ , \qquad \ 4\Gamma v_{\mbox{\tiny max}} + 4\xi^{2} \ \ll \ \Omega^{2} \ .
\end{eqnarray}
Here we used (\ref{MaximizadorfRWA}) to evaluate \ $f_{\mbox{\tiny RWA}}(x_{2n})$ \ for \ $\xi \gg 1$. 

We now illustrate (\ref{ValidezFriccionSinRWA2}) for the parameters of Fig.~\ref{Figure10}. The figure has \ $\xi = 10$, \ $v_{\mbox{\tiny max}} \leq 2$, \ $100 \leq \Omega$, and \ $\Gamma = 0.07$. Hence, \ $2v_{\mbox{\tiny max}} \leq 4  \ll 100 \leq \Omega$ \ and \ $4\Gamma v_{\mbox{\tiny max}} + 4\xi^{2} \leq 400.56 \ll 10^{4} \leq \Omega^{2}$, so that (\ref{1condicionesEq}) is satisfied for \ $t \geq 0$. Now we consider the case of 
Fig.~\ref{Figure11}. Figure \ref{Figure11a} has \ $\xi = 10$, \ $\Omega = 6$, \ $\Gamma = 1$, \ $\vert x'(\tau) \vert \leq 0.26$, \ and \ $\vert x''(\tau) \vert \leq 1.8$ \ for \ $\tau \geq 0$. Using (\ref{CorrespondenciaX}) it follows that \ $\vert \ddot{q}(t)/(c\omega_{0}) \vert \leq 0.2$ \ and \ $\vert \dot{q}(t)/c \vert \leq 0.087$ \ for \ $t\geq 0$. Therefore, (\ref{1condicionesEq}) is not satisfied and Fig.~\ref{Figure11a} corresponds to a non-physical solution of the model (terms of order $\dot{q}(t)/c$ and $\ddot{q}(t)/(c\omega_{0})$ would have to be introduced). On the other hand, figure \ref{Figure11b} has \ $\xi = 50$, \ $\Omega = 14.1$, \ $\Gamma = 1$, \ $\vert x'(\tau) \vert \leq 0.1 $, \ and \ $\vert x''(\tau) \vert \leq 1.5$ \ for \ $\tau \geq 15$. Using (\ref{CorrespondenciaX}) it follows that \ $\vert \ddot{q}(t)/(c\omega_{0}) \vert \leq 0.03$ \ and \ $\vert \dot{q}(t)/c \vert \leq 0.015$ \ for \ $\tau = \Delta t \geq 15$. Therefore, (\ref{1condicionesEq}) is satisfied for $\tau \geq 15$. Nevertheless, (\ref{1condicionesEq}) is not satisfied for small $\tau$ because the mirror is subject to large accelerations (it has \ $\vert x''(\tau) \vert \leq 460$ \ for \ $\tau \geq 0$). Hence, the model used is only applicable for the steady-state solution in Fig.~\ref{Figure11b}.

Finally we note that (\ref{91}) can be used to solve (\ref{F1}) approximately if \ $\xi \gg 1$. We do not pursue this direction here.


\section{INTRODUCTION OF A HARMONIC OSCILLATOR POTENTIAL}

In the previous section it was shown that, in the RWA regime and for \ $\xi \gtrsim 5$, the addition of friction  leads to a final state in which the mirror is at rest at a position where $\omega_{0}$ is very different from one of the cavity resonance frequencies. In experimental set-ups the movable mirror normally executes small oscillations around a position where $\omega_{0}$ does coincide with one of the cavity resonance frequencies. In order to achieve this we assume in this section that the movable mirror is confined by a harmonic oscillator potential. The equation governing the dynamics of the mirror is
\begin{eqnarray}
\label{HO12}
&& \frac{d^{2}x}{d\tau^{2}}(\tau) + \Gamma \frac{dx}{d\tau}(\tau) + \omega_{\mbox{\tiny ho}}^{2}\left[ x(\tau ) - x_{\mbox{\tiny E}} \right] \cr
&& \cr
&=& \left\{ 1 + \mbox{cos}\left[ \Omega\tau + 2x(\tau) -2\delta_{k_{\scriptscriptstyle{N}}^{\scriptscriptstyle{0}}}\left( \frac{x(\tau)}{k_{\scriptscriptstyle{N}}^{\scriptscriptstyle{0}}} \right) \right] \right\} \times \cr
&& \qquad \times f_{\mbox{\tiny RWA}}\left[ x(\tau) \right] \ . 
\end{eqnarray}
Equation (\ref{HO12}) is obtained by adding on the right-hand side of (\ref{2}) the force \ $-\gamma \dot{q}(t) - k_{\mbox{\tiny ho}}[q(t)-q_{\mbox{\tiny E}}]$ \ and then using (\ref{76}) to express the equation in terms of non-dimensional quantities. Here $\gamma$ has units of Ns/m$^{3}$ and $k_{\mbox{\tiny ho}}$ has units of N/m$^{3}$. Also, \ $\Gamma = \gamma/(\Delta M_{0})$, \ $\omega_{\mbox{\tiny ho}} = \sqrt{k_{\mbox{\tiny ho}}/(\Delta^{2}M_{0})}$, and \ $x_{\mbox{\tiny E}} = k_{\mbox{\tiny N}}^{\mbox{\tiny 0}} q_{\mbox{\tiny E}}$ \ are non-dimensional quantities. 

In the rest of this section we assume that \ $x_{\mbox{\tiny E}} = x_{2n}$ \ for some \ $n\in\mathbb{Z}^{+}$, so that the harmonic oscillator potential is centred at a position where $\omega_{0}$ coincides with one of the cavity resonance frequencies, see (\ref{ExtraMaximizadoresQ2n}) and (\ref{MaximizadorfRWA}) and the definition of $\xi$ in (\ref{76}). 

\subsection{The limit of low field intensity}

The limit of low field intensity is still defined by condition (\ref{CondicionIntensidadDebil}). It allows us to eliminate the explicit time dependence in (\ref{HO12}) to obtain
\begin{eqnarray}
\label{HO12rwa}
\frac{d^{2}x}{d\tau^{2}}(\tau) + \Gamma \frac{dx}{d\tau}(\tau) + \omega_{\mbox{\tiny ho}}^{2}\left[ x(\tau) -x_{\mbox{\tiny E}} \right] &=& f_{\mbox{\tiny RWA}}\left[ x(\tau) \right]  \ . \cr
&&
\end{eqnarray}
Notice that (\ref{HO12rwa}) can be written as
\begin{eqnarray}
\label{HO12rwaBIS}
\frac{d^{2}x}{d\tau^{2}}(\tau) + \Gamma \frac{dx}{d\tau}(\tau) &=& -\frac{d\mathcal{V}}{dx}\left[ x(\tau) \right]  \ ,  \end{eqnarray}
where $\mathcal{V}(x)$ is the potential affecting the motion of the mirror. It is given by 
\begin{eqnarray}
\label{ExtraPotencialHO}
\mathcal{V}(x) &=& V_{\mbox{\tiny RWA}}(x) + \frac{1}{2}\omega_{\mbox{\tiny ho}}^{2}(x-x_{\mbox{\tiny E}})^{2} \ .
\end{eqnarray}

We now perform a phase space analysis of (\ref{HO12rwa}). The first order non-linear system associated with (\ref{HO12rwa}) is
\begin{eqnarray}
\label{HOrwa2}
\frac{d}{d\tau}\left(
\begin{matrix}
x(\tau) \cr
y(\tau)
\end{matrix}
\right)
&=&
\mathbf{f}_{\mbox{\tiny ho}}\left[ x(\tau), y(\tau) \right] \ .
\end{eqnarray}
Here \ $y(\tau) = x'(\tau)$ \ and
\begin{eqnarray}
\mathbf{f}_{\mbox{\tiny ho}}(x,y) \ = \ 
\left(
\begin{matrix}
y \cr
-\Gamma y - \omega_{\mbox{\tiny ho}}^{2}(x-x_{\mbox{\tiny E}}) + f_{\mbox{\tiny RWA}}(x) 
\end{matrix}
\right) \ .
\end{eqnarray}

Since \ $\nabla \cdot \mathbf{f}_{\mbox{\tiny ho}}(x,y) = - \Gamma < 0$ \ for all \ $x,y \in\mathbb{R}$, Bendixon's criterion (Theorem 1 in Section 3.9 of \cite{Strogatz}) allows us to conclude that there are no closed orbits and no cycle-graphs in phase space. Moreover, the fixed points \ $(x_{n}^{\mbox{\tiny ho}}, y_{n}^{\mbox{\tiny ho}})$ \ of (\ref{HOrwa2}) are defined by
\begin{eqnarray}
\label{HOrwa4}
y_{n}^{\mbox{\tiny ho}} = 0 \ , \qquad f_{\mbox{\tiny RWA}}(x_{n}^{\mbox{\tiny ho}}) = \omega_{\mbox{\tiny ho}}^{2}(x_{n}^{\mbox{\tiny ho}} - x_{\mbox{\tiny E}}) \ \ .
\end{eqnarray}
Notice that $x_{n}^{\mbox{\tiny ho}}$ is a critical point of the potential $\mathcal{V}(x)$, since the second equation in (\ref{HOrwa4}) can be written as \ $(d\mathcal{V}/dx)(x_{n}^{\mbox{\tiny ho}}) = 0$. 

In the following we take $x_{n}^{\mbox{\tiny ho}}$ to be the first (an possibly only) solution of the second equation in (\ref{HOrwa4}) located to the right of the maximizer \ $x_{\mbox{\tiny E}} = x_{2n}$ \ of $f_{\mbox{\tiny RWA}}(x)$ for each \ $n\in\mathbb{Z}^{+}$. Then $x_{n}^{\mbox{\tiny ho}}$ is a minimizer of $\mathcal{V}(x)$ (this conclusion is immediate if one observes the graph of $V_{\mbox{\tiny RWA}}(x)$ in Fig.~\ref{Figure3a} and adds a harmonic oscillator potential centred at \ $x= n\pi + 1/\xi$). From (\ref{MinimizadorVRWA}) and (\ref{MaximizadorfRWA}) it follows that $x_{n}^{\mbox{\tiny ho}}$ is located between a maximizer $x_{2n}$ of $f_{\mbox{\tiny RWA}}(x)$ and a minimizer $x_{n}^{**}$ of $V_{\mbox{\tiny RWA}}(x)$ if \ $\xi \gg 1$, that is, \ $x_{2n} \simeq n\pi + 1/\xi < x_{n}^{\mbox{\tiny ho}} < n\pi + 2/\xi \simeq x_{n}^{**}$ \ if \ $\xi \gg 1$.

The equation for \ $x_{n}^{\mbox{\tiny ho}}$ \ in (\ref{HOrwa4}) can be solved to good approximation if \ $\xi \gg 1$ \ and one uses (\ref{MaximizadorfRWA}) and the displaced Lorentzian approximation of $f_{\mbox{\tiny RWA}}(x)$ given in (\ref{ExtraLorentziana}). One has to solve the cubic equation
\begin{equation}
\label{hoP6}
u_{n}^{3} + \frac{1}{2\omega_{\mbox{\tiny ho}}^{2}}u_{n}^{2} + \frac{1}{\xi^{4}}u_{n} \ = \ \frac{1}{2\xi^{2}\omega_{\mbox{\tiny ho}}^{2}}\left( 1 - \frac{1}{\xi^{2}} \right) \ ,
\end{equation}
with \ $u_{n} = x_{n}^{\mbox{\tiny ho}} - (n\pi +1/\xi)$. Using Descartes' rule of signs \cite{Algebra} it follows that (\ref{hoP6}) has exactly one positive real root (recall that we have defined $x_{n}^{\mbox{\tiny ho}}$ to satisfy \ $x_{n}^{\mbox{\tiny ho}} > x_{\mbox{\tiny E}}$ \ so that \ $u_{n} >0$). It can be calculated explicitly (using the Cardan's formulas \cite{Algebra} or symbolic calculations), but the expression is quite long. For this reason we have decided to omit it.

A linear stability analysis of (\ref{HOrwa2}) shows that the fixed point \ $(x_{n}^{\mbox{\tiny ho}}, y_{n}^{\mbox{\tiny ho}})$ \ is always an attractor. To prove this result one has to use that \ $f_{\mbox{\tiny RWA}}'(x_{n}^{\mbox{\tiny ho}}) < 0$, since $f_{\mbox{\tiny RWA}}(x)$ is strictly decreasing from its maximizer $x_{2n}$ to $x_{n}^{**}$ where it is zero and $x_{n}^{\mbox{\tiny ho}}$ is located between them (see items 6 and 8 on page 6). In particular, \ $(x_{n}^{\mbox{\tiny ho}}, y_{n}^{\mbox{\tiny ho}})$ \ is a stable spiral (Theorem 4 of Section 2.10 of \cite{Strogatz}) if
\begin{eqnarray}
\label{HOrwa3}
\left(\frac{\Gamma}{2}\right)^{2} \ < \ \omega_{\mbox{\tiny ho}}^{2} - f_{\mbox{\tiny RWA}}'(x_{n}^{\mbox{\tiny ho}}) \ .
\end{eqnarray}
On the other hand, it is a stable node (Theorem 4 of Section 2.10 of \cite{Strogatz}) if the $<$ is replaced by $\geq$ in (\ref{HOrwa3}). Notice that a typical value is \ $\Gamma /\omega_{\mbox{\tiny ho}} \sim 10^{-3}$ \ (see Appendix III), which gives a fixed point that is a stable spiral. 

In general, the movable mirror tends to a minimizer or maximizer of $\mathcal{V}(x)$ as \ $\tau \rightarrow +\infty$. This follows from the Poincar\'{e}-Bendixon theory in $\mathbb{R}^{2}$ and from the facts that the energy \ $|x'(\tau)|^{2}/2 + \mathcal{V}(x)$ \ of the movable mirror is always decreasing and that there are no closed orbits and cycle-graphs in phase space.

The approximation of $V_{\mbox{\tiny RWA}}(x)$ in (\ref{91}) can be used to obtain an analytic approximation of the solution of (\ref{HO12rwa}), but this analytic approximation is not accurate. The reason is that the evolution predicted by the approximate $x(\tau)$ is qualitatively different and tends to $x_{\mbox{\tiny E}} = x_{2n}$ instead of $x_{n}^{\mbox{\tiny ho}}$ if \ $\tau \rightarrow + \infty$. The displaced Lorentzian approximation for $f_{\mbox{\tiny RWA}}(x)$ could also be used, but it seems that there is no explicit perturbative solution available (the perturbation parameter being $1/\xi$). Instead, we shall give an analytic approximation that qualitatively describes the solution in the general case of (\ref{HO12}) in the next subsection. 

We now establish sufficient conditions for (\ref{1condicionesEq}) to be satisfied. We first consider the condition on the velocity of the mirror. From (\ref{HO12rwaBIS}) it follows that the (non-dimensional) energy of the mirror is given by
\begin{eqnarray}
\label{energiaHO}
E_{\mbox{\tiny ho}}[x(\tau),x'(\tau)] &=& \frac{1}{2}\left[ x'(\tau) \right]^{2} + \mathcal{V}\left[ x(\tau) \right] \  ,
\end{eqnarray}
and that it is a decreasing function of time
\begin{eqnarray}
\label{energiaHO2}
E_{\mbox{\tiny ho}}[x(\tau),x'(\tau)] &=& E_{\mbox{\tiny ho}}^{0} -  \Gamma\int_{0}^{\tau} d\tau' \left[ x'(\tau') \right]^{2} \ , 
\end{eqnarray}
with \ $E_{\mbox{\tiny ho}}^{0} \equiv E_{\mbox{\tiny ho}}[x(0),x'(0)]$ \ and \ $\tau \geq 0$. Hence, one can bound the velocity of the mirror using the bounds for $V_{\mbox{\tiny RWA}}(x)$ in (\ref{CotasVRWA}) and (\ref{CorrespondenciaX}), (\ref{energiaHO}), and (\ref{energiaHO2}). One obtains
\begin{eqnarray}
\label{ValidezHO1}
\left\vert \frac{\dot{q}(t)}{c} \right\vert  
&\leq& \frac{2}{\Omega}\sqrt{2E_{\mbox{\tiny ho}}^{0} + \pi} \qquad (t \geq 0). \ \ 
\end{eqnarray}
Again, one essentially needs \ $\Omega \gg 1$. 

We now consider the the condition on the acceleration of the mirror. Using the triangle inequality and (\ref{MaximizadorfRWA}), it follows from (\ref{CorrespondenciaX}) and (\ref{HO12rwa}) that
\begin{eqnarray}
\label{ValidezHORWApre2}
\left\vert \frac{\ddot{q}(t)}{c\omega_{0}} \right\vert &\leq& \frac{8}{\Omega^{2}}f_{\mbox{\tiny RWA}}(x_{2n}) + \frac{2\Gamma}{\Omega} \left\vert \frac{\dot{q}(t)}{c} \right\vert \cr
&& \ \  + \frac{4\omega_{\mbox{\tiny ho}}^{2}}{\Omega^{2}}\left\vert x(\tau)-x_{\mbox{\tiny E}} \right\vert \ , 
\end{eqnarray}
with \ $t\in\mathbb{R}$ \ and \ $\tau = \Delta t$. Again, the second condition in (\ref{1condicionesEq}) will also be satisfied if $\Omega$ is large enough. 

In particular, if \ $\xi \gg 1$ \ (say, \ $\xi \gtrsim 5$) and the movable mirror is confined to the interval \ $(n\pi , \ n\pi + 2/\xi )$ \ for \ $\tau \geq 0$, then it follows from (\ref{ValidezHO1}) and (\ref{ValidezHORWApre2}) that (\ref{1condicionesEq}) is satisfied if 
\begin{eqnarray}
\label{OtroValidezHO}
\Gamma , \ \sqrt{2E_{\mbox{\tiny ho}}^{0} + \pi}, \ \xi, \ \omega_{\mbox{\tiny ho}} &\ll& \frac{\Omega}{2} \ . 
\end{eqnarray}
Notice that typical values are \ $\omega_{\mbox{\tiny ho}}/\Omega \sim 10^{-11}$, \ $\Gamma/\Omega \sim 10^{-14}$, \ $\xi/\Omega \sim 10^{-8}$, \ and \ $\Omega \sim 10^{9}$, see Appendix III. Also, the end points of the interval \ $(n\pi , n\pi + 2/\xi)$ \ are (approximately) consecutive maximizers and minimizers of $V_{\mbox{\tiny RWA}}(x)$ and the non-dimensional position \ $x_{\mbox{\tiny E}} = x_{2n}$ \ where $\omega_{0}$ coincides with one of the cavity resonance frequencies is located between them, see (\ref{MaximizadorVRWA})-(\ref{MaximizadorfRWA}).

Finally, we discuss the validity of the RWA. In the following subsection it is shown that, after a transient, the solution of (\ref{HO12}) approximately describes an oscillation of frequency $\Omega$ around the fixed point $x_{n}^{\mbox{\tiny ho}}$ whose amplitude is estimated by 
\begin{eqnarray}
\label{Amplitud}
\frac{\xi^{2}-1}{2\sqrt{\Gamma^{2}\Omega^{2} + (\Omega^{2}-\omega_{\mbox{\tiny ho}}^{2})^{2}}} \ .
\end{eqnarray}
Hence, the amplitude of the oscillation will be small if
\begin{eqnarray}
\label{HOCondicionRWA}
\frac{\xi^{2}-1}{2} &\ll& \sqrt{\Gamma^{2}\Omega^{2} + (\Omega^{2}-\omega_{\mbox{\tiny ho}}^{2})^{2}} \ \simeq \Omega^{2} \ .
\end{eqnarray}
In the approximate equality in (\ref{HOCondicionRWA}) we have used that one typically has \ $\omega_{\mbox{\tiny ho}}/\Omega \ll 1$ \ and \ $\Gamma/\Omega \ll 1$ \ (see Appendix III). We take (\ref{HOCondicionRWA}) as a condition necessary for the RWA to be valid, since the amplitude of the steady state oscillations will be very small and the movable mirror will be found approximately at the fixed point $x_{n}^{\mbox{\tiny ho}}$ just as predicted in the RWA regime. Notice that the RWA regime is compatible with the conditions in (\ref{1condicionesEq}), since both require $\Omega$ sufficiently large.

\subsection{The intermediate case}

In this subsection we treat the general case in (\ref{HO12}), where the movable mirror does not have conservative dynamics because the force affecting it depends explicitly on time. In all that follows we take
\begin{equation}
\label{UU}
u(\tau) \ = \ x(\tau) -x_{\mbox{\tiny E}}\ , \qquad u_{n} \ = \ x_{n}^{\mbox{\tiny ho}} -x_{\mbox{\tiny E}} \ .
\end{equation}
We first give an analytic approximation of the steady-state solution in the case where the transparency of the movable mirror is small (that is, $\xi \gg 1$).

First observe that friction and the harmonic oscillator potential bound the motion of the mirror to a finite number of wells of $V_{\mbox{\tiny RWA}}(x)$, that is, for all \ $\tau\in\mathbb{R}$ \ one has
\begin{eqnarray}
\label{Despreciar80p1}
n\pi \ \leq \ x(\tau) \ \leq \ (n+m)\pi \ \ \mbox{for some} \ \ n < m\in\mathbb{Z}^{+} \ .
\end{eqnarray}
Then one can take \ $\delta_{k_{\scriptscriptstyle{N}}^{\scriptscriptstyle{0}}}[q(t)] \in [-\pi, m\pi]$ \ and (\ref{HO12}) reduces to
\begin{eqnarray}
\label{despreciar80}
&& x''(\tau) + \Gamma x'(\tau) + \omega_{\mbox{\tiny ho}}^{2}[x(\tau) - x_{\mbox{\tiny E}}] \cr
&\simeq& \left[ 1 +\mbox{cos}(\Omega \tau) \right] f_{\mbox{\tiny RWA}}[x(\tau)] \ , \qquad 
\end{eqnarray} 
for \ $\tau \gg 1$. More precisely, one needs
\begin{eqnarray}
\label{despreciar80p2}
\tau \ \gg \ \frac{2\pi}{\Omega}(n+m) \ .
\end{eqnarray}

Assume that \ $\xi \gg 1$, \ $\vert \xi^{2}u(\tau) \vert \ll 1$ \ for \ $\tau \geq 0$, and that $\tau$ satisfies (\ref{despreciar80p2}). Then (\ref{despreciar80}) can be approximated as
\begin{eqnarray}
\label{hoP10}
u''(\tau) + \Gamma u'(\tau) + \omega_{\mbox{\tiny ho}}^{2}u(\tau) &\simeq& \frac{\xi^{2}-1}{2}\left[ 1 + \mbox{cos}(\Omega \tau)\right] \ . \cr
&&  
\end{eqnarray} 
Here we used the displaced Lorentzian approximation to $f_{\mbox{\tiny RWA}}(x)$ in (\ref{ExtraLorentziana}) and the assumptions \ $\xi \gg 1$ \ and \ $\vert \xi^{2}u(\tau) \vert \ll 1$ \ $(\tau \geq 0)$ \ as follows 
\begin{eqnarray}
\label{hoP12}
f_{\mbox{\tiny RWA}}\left[ u(\tau) + x_{\mbox{\tiny E}} \right] &\simeq& -\frac{1}{2} + \frac{\xi^{2}}{2}\frac{1}{[\xi^{2}u(\tau)]^{2} + 1} \ , \cr
&& \cr
&\simeq& -\frac{1}{2} + \frac{\xi^{2}}{2} \qquad (\tau \geq 0).
\end{eqnarray}

Equation (\ref{hoP10}) corresponds to a forced harmonic oscillator with damping. The transient term decays exponentially to zero as \ $\tau \rightarrow + \infty$ \ and the steady-state solution $u_{\mbox{\tiny st}}(\tau)$ is given by 
\begin{eqnarray}
\label{hoP15}
u_{\mbox{\tiny st}} (\tau) &=& \frac{\xi^{2}-1}{2\omega_{\mbox{\tiny ho}}^{2}} + \frac{\xi^{2}-1}{2}\frac{\mbox{cos}(\Omega\tau - \theta)}{\sqrt{\Gamma^{2}\Omega^{2} + (\Omega^{2}-\omega_{\mbox{\tiny ho}}^{2})^{2}}} \ ,\cr
&& \cr
\theta &=& \mbox{cos}^{-1}\left[ \frac{\omega_{\mbox{\tiny ho}}^{2}-\Omega^{2}}{\sqrt{\Gamma^{2}\Omega^{2} + (\Omega^{2}-\omega_{\mbox{\tiny ho}}^{2})^{2}}} \right] \in \left(0,\pi \right) \ . \cr
&&
\end{eqnarray}
Although the conditions under which (\ref{hoP15}) was derived are rather stringent, $u_{\mbox{\tiny st}}(\tau)$ qualitatively describes rather well the exact solution. The steady-state exact solution behaves asymptotically as  \ $u(\tau) \simeq u_{0} + A\mbox{cos}(\Omega\tau - \psi)$, where \ $u_{0} \simeq u_{n}$ given in (\ref{UU}). Therefore, (\ref{hoP15}) identifies both the frequency $\Omega$ of the steady-state oscillations of the mirror and the fact that resonance in the amplitude of the steady-state oscillations of the mirror occurs when $\Omega = \omega_{\mbox{\tiny ho}}$. Hence, resonance occurs if the angular frequency \ $\Delta\omega_{\mbox{\tiny ho}}$ \ of the harmonic oscillator potential is twice the angular frequency of the field. Note that one is normally out of resonance, since one typically has \ $2\omega_{0} \gg \Delta\omega_{\mbox{\tiny ho}}$ \ (from Appendix III a typical value is \ $\omega_{\mbox{\tiny ho}}/\Omega \sim 10^{-11}$ \ for field frequencies in the optical regime).  We now illustrate these remarks through numerical calculations. 

\begin{figure}[htbp]
  \centering
  \subfloat[]{\label{Figure12a}\includegraphics[scale=0.75]{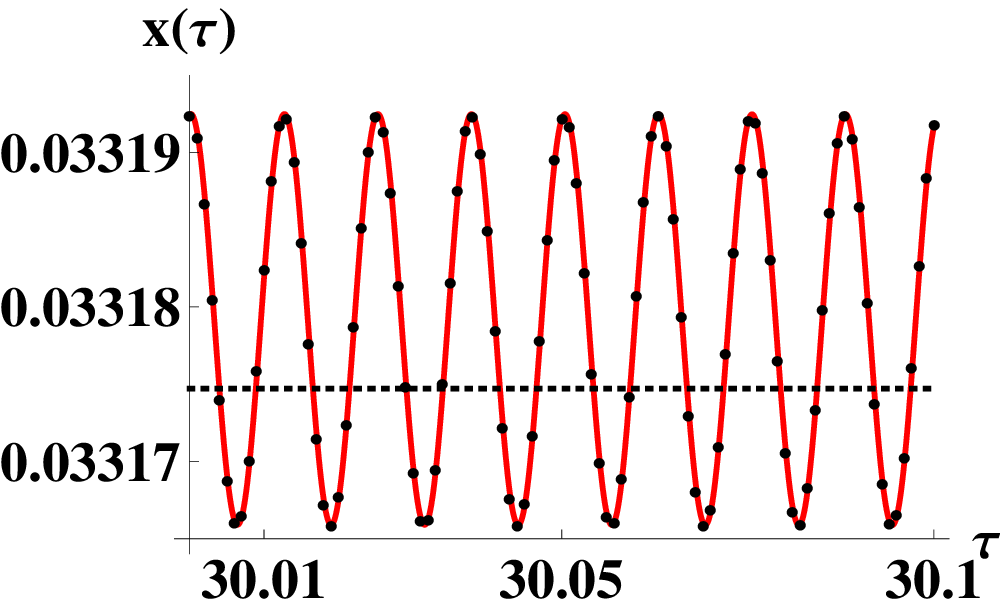}} \hspace{0.3cm}
  \subfloat[]{\label{Figure12b}\includegraphics[scale=0.75]{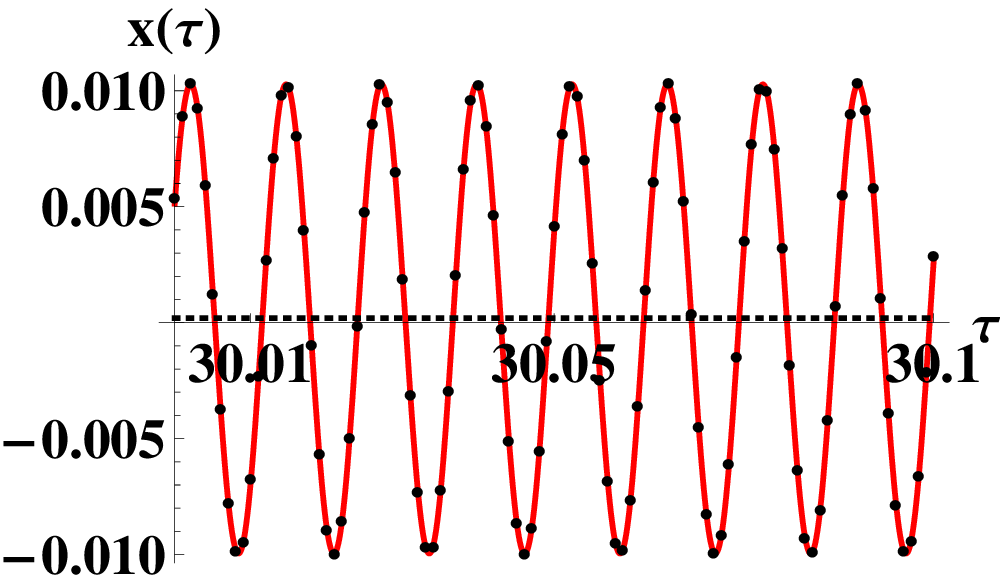}} \\
  \caption{(Color online) The figures show \ $u(\tau) = x(\tau) -x_{\mbox{\tiny E}}$ \ (red-solid line) calculated numerically from (\ref{HO12}), the function \ $u_{0} + A\mbox{cos}(\Omega\tau - \psi)$ \ (black-dotted line), and $u_{n} = x_{n}^{\mbox{\tiny ho}} -x_{\scriptscriptstyle{E}} $ (horizontal black-dotted line) for \ $\xi=10$, \ $\Gamma = 1$, and \ $\Omega = 500$. Figure \ref{Figure12a} has \ $\omega_{\mbox{\tiny ho}} = 10$, \ $A =1.33\times 10^{-5}$, \ $\psi = (\pi +1)/2$, \ $u_{0} = u_{n} +4.4\times 10^{-6}$, and \ $u_{n} = 0.033175$. Figure \ref{Figure12b} has \ $\omega_{\mbox{\tiny ho}} = 500$, \ $A = 10^{-2}$, \ $\psi = (\pi +3)/2$, \ $u_{0} = u_{n} = 2 \times 10^{-4}$.}
\label{Figure12}
\end{figure}

Figure~\ref{Figure12a} illustrates \ $u(\tau)$ \ (red-solid line) calculated numerically from (\ref{HO12}) once the transient term is negligible and compares it with the function \ $u_{0} + A\mbox{cos}(\Omega\tau - \psi)$ \ (black-dotted line) with \ $A = 1.33 \times 10^{-5}$, \ $\psi = (\pi+1)/2$ \ and \ $u_{0} = u_{n} + 4.4 \times 10^{-6}$ \ with $u_{n} = 0.033175$ \ in (\ref{UU}). The parameters are \ $\xi = 10$, \ $\Omega = 500$, \ $\omega_{\mbox{\tiny ho}} = 10$, and \ $\Gamma = 1$.  Notice that the system is far off resonance (that is, $\Omega \gg \omega_{\mbox{\tiny ho}}$). Hence, the movable mirror executes very small oscillations around $u_{n}$, since the steady-state oscillations have an amplitude approximately equal to \ $1.33\times 10^{-5}\lambda/(2\pi)$ \ with $\lambda$ the wavelength of the field. Moreover, the mirror oscillates out of the region \ $[- \xi^{-2} = -0.01, \ \xi^{-2} = 0.01]$ \ where $\omega_{0}$ coincides with one of the cavity resonance frequencies. 

Figure~\ref{Figure12b} illustrates \ $u(\tau)$ \ (red-solid line) calculated numerically from (\ref{HO12}) and compares it with the function \ $u_{n} + A\mbox{cos}(\Omega\tau - \psi)$ \ (black-dotted line) with \ $u_{n} = 2\times 10^{-4}$ \ in (\ref{UU}), \ $A = 10^{-2}$, and \ $\psi = (\pi + 3)/2$. The parameters are \ $\xi = 10$, \ $\Omega = \omega_{\mbox{\tiny ho}} = 500$, \ and \ $\Gamma = 1$. Observe that the movable mirror oscillates in the region \ $[-1/\xi^{2} = -10^{-2}, \ 1/\xi^{2} = 10^{2}]$ \ where $\omega_{0}$ coincides with a cavity resonance frequency. Introducing units it follows that the steady-state oscillations of the mirror have approximately an amplitude of \ $10^{-2}\lambda/(2\pi)$.

To end this subsection we give sufficient conditions for the validity of (\ref{1condicionesEq}). In a similar way as in the previous subsection it can be shown from (\ref{HO12}) that (\ref{ValidezHORWApre2}) still holds. The problem is that there is no direct way to bound the velocity of the mirror, since energy is no longer conserved. Nevertheless, using the approximate $u(\tau)$ in (\ref{hoP15}) in combination with (\ref{CorrespondenciaX}) one can obtain the following estimate:
\begin{eqnarray}
\label{NuevoValidezHO}
\left\vert \frac{\dot{q}(t)}{c} \right\vert &\lesssim& \frac{\vert \xi^{2}-1 \vert}{\sqrt{\Gamma^{2}\Omega^{2} + (\Omega^{2}-\omega_{\mbox{\tiny ho}}^{2})^{2}}} \qquad (\tau \gg 1).
\end{eqnarray}
From (\ref{ValidezHORWApre2}) and (\ref{NuevoValidezHO}) it follows that (\ref{1condicionesEq}) will be satisfied if, for example, \ $\Omega \gg \xi, \ \Gamma , \ \omega_{\mbox{\tiny ho}}$ \ so that the field is far off resonance with the frequency of the harmonic oscillator potential. For the typical values in Appendix III one has \ $\Gamma/\Omega \sim 10^{-14}$ \ and \ $\omega_{\mbox{\tiny ho}}/\Omega \sim 10^{-11}$ \ and \ $\xi/\Omega \sim 10^{-8} $. Also, the parameters of Fig.~\ref{Figure12a} satisfy this so the model is applicable in that case.


\section{CONCLUSIONS}

In this article we considered one of the paradigmatic models in optomechanics: a one-dimensional cavity composed of a perfect, fixed mirror and of a movable mirror with non-zero transparency. Moreover, we considered the following physical situation: on the far right a monochromatic laser of frequency $\omega_{0}$ is always turned on. The plane wave associated with the laser travels to the left, is partially reflected by the movable mirror at \ $q(t)>0$, and completely reflected by the perfect mirror fixed at \ $x=0$. After a transient, a standing wave is formed and the laser is responsible for maintaining it. We described the dynamics of the system after the aforementioned transient. To establish the equations of the system we neglected terms of order $\dot{q}(t)/c$ and $\ddot{q}(t)/(c\omega_{0})$, that is, we only considered the leading term in the force affecting the movable mirror. As a consequence of these approximations, the single-mode approximation is possible and the friction force affecting the movable mirror and associated with the interaction with the field does not appear \cite{Nuestro, DomokosI}. Since we consider only the case where \ $\vert \dot{q}(t)/c \vert, \ \vert \ddot{q}(t)/(c\omega_{0})\vert \ll 1$, the friction force and mixing of field amplitudes corresponding to different wave-numbers are very small.  

Our approach differs from most of the articles in that the mirror can deviate far from an equilibrium position and we considered the exact modes of the whole system. This allowed us to conclude that in the rotating-wave-approximation (RWA) the radiation pressure force is derived from a periodic potential with period half the wavelength of the field and that the dynamics of the mirror due to radiation pressure are determined by the mirror approaching or withdrawing from a \textit{resonance position} where the frequency of the field coincides with one of the cavity resonance frequencies. Moreover, there is only one resonance position in each well of the potential. 

Special attention was given to establishing and verifying conditions that determine when the equations used to describe the physical system are accurate approximations to the exact Maxwell-Newton equations. These amount to asking that the velocity and acceleration of the mirror be small enough so that the electromagnetic field evolves as if the mirror were instantaneously fixed. We identified three regimes: the low intensity or RWA regime, the high intensity regime, and the intermediate regime. The first one occurs when the frequency of the field is much larger than $1$ over the characteristic time-scale of the movable mirror, the second when the former is much smaller than the latter, and the third treats the values in between. We found that the RWA regime is a natural setting for the equations, while the intermediate regime is compatible with them for a certain range of parameters. The high intensity regime is incompatible with the equations used, since the movable mirror can be subject to large accelerations in this case. In order to treat the high intensity regime and the intermediate case in some range of parameters it is fundamental to include the corrections of order $\dot{q}(t)/c$ and $\ddot{q}(t)/(c\omega_{0})$ to the force affecting the movable mirror. Therefore, effects such as dissipation and mixing of field amplitudes corresponding to different wave-numbers \cite{Nuestro,DomokosI} become relevant in this case. Also, it is important to note that the system exhibits conservative dynamics only in the RWA regime. 

The introduction of the radiation pressure potential gave rise to an intuitive physical understanding of the dynamics of the movable mirror. In the RWA regime and when only radiation pressure is considered, the movable mirror has constant energy and can have either a bounded or unbounded trajectory where it accelerates as it approaches a minimizer and decelerates as it approaches a maximizer of the radiation pressure potential. In the case of a bounded trajectory, the movable mirror is bound to one of the potential wells and executes non-harmonic oscillations around one of the minimizers of the radiation pressure potential. The consideration of the exact modes of the whole system allowed us to explain this trajectory as follows. As the mirror approaches the resonance position, the cavity field increases and exerts a radiation pressure that surpasses the one exerted by the field outside the cavity and that pushes the mirror away from the resonance position. As the mirror withdraws from the resonance position, the cavity field is emptied and the field outside exerts a larger force that pushes the mirror back to the resonance position. This process repeats itself over and over giving rise to the periodic trajectory of the mirror. Outside the RWA regime the dynamics of the mirror are much more complex since the mirror can jump between potential wells.  

In the RWA regime and when the movable mirror is also subject to friction linear in the velocity, the movable mirror continually losses energy and eventually tends to the position of a minimizer (or maximizer) of the radiation pressure potential. Since this position is very different from the resonance position, the cavity field is emptied. Outside the RWA regime the dynamics are much more complex, since the movable mirror can jump between wells of the radiation pressure potential before settling in one of its minimizers (or maximizers). For some values of the parameters there is numerical evidence that the mirror can even execute periodic oscillations around one the minimizers. 

In the RWA regime and when the movable mirror is subject both to friction and to a harmonic oscillator potential centred at a resonance position, the movable mirror always tends to a minimizer (or maximizer) of the sum of the harmonic oscillator and radiation pressure potentials. Outside the RWA regime and after a transient, the movable mirror settles in a harmonic oscillation of frequency 2 times the frequency of the field and around an aforementioned minimizer. Moreover, there is resonance in the amplitude if the frequency of the harmonic oscillator is equal to two times the frequency of the field.  

In future work we will include in the dynamics the friction force associated with the interaction with the field.

\section*{ACKNOWLEDGEMENTS}

We thank Pablo Barberis-Blostein, Marc Bienert, and Christian Gerard for fruitful discussions. L. O. Casta\~{n}os
wishes to thank the Universidad Nacional Aut\'{o}noma de M\'{e}xico for support. Research partially supported by CONACYT under Project CB-2008-01-99100. We thank the referee for helpful suggestions that helped improve the presentation of the results.

\begin{enumerate}
\item[*] E-mail: LOCCJ@yahoo.com
\item[+] Fellow, Sistema Nacional de Investigadores. E-mail: weder@unam.mx
\end{enumerate}


\section*{APPENDIX I}

In this appendix we calculate the maximizers and minimizers of $L_{k}(q)$ given in (\ref{7Lk}) as a function of \ $q \geq 0$ \ for fixed \ $k >0$ \ and \ $\chi_{0} >0$. 

First observe from the definition of $L_{k}(q)$ in (\ref{7Lk}) that
\begin{eqnarray}
\label{B1}
L_{k}(q) &=& \sqrt{\frac{2}{\pi}}\left[ 1+\frac{( 4\pi\chi_{0}k )^{2}}{2} -(4\pi\chi_{0}k)\mbox{sin}(2kq) \right. \cr
&& \ \ \ \qquad \left. -\frac{(4\pi\chi_{0}k)^{2}}{2}\mbox{cos}(2kq) \right]^{-1/2} \ .
\end{eqnarray}
It then follows that
\begin{eqnarray}
\label{B3}
\frac{dL_{k}}{dq}(q_{n}) \ = \ 0 &\Leftrightarrow& \mbox{tan}(2kq_{n}) = \frac{2}{4\pi\chi_{0}k} \ .
\end{eqnarray}

In the following assume that
\begin{eqnarray}
\label{i1bis}
4\pi\chi_{0}k \ \gg \ 1 \ .
\end{eqnarray}
From (\ref{NuevoTransparencia}) it follows that (\ref{i1bis}) implies that the transparency of the movable mirror is very small.

Using (\ref{i1bis}) it follows that a first approximation \ $q_{n} \simeq n\pi/(2k)$ \ with \ $n\in \mathbb{Z}^{+}$ \ ($\mathbb{Z}^{+}$ is the set of non-negative integers) is obtained by neglecting the term $2/(4\pi\chi_{0}k)$ on the right of (\ref{B3}). Using Newton's method \cite{Burden} one can obtain a better approximation of $q_{n}$. After one iteration it follows that
\begin{eqnarray}
\label{B5}
q_{n} &\simeq& \frac{1}{2k}\left( n\pi + \frac{2}{4\pi\chi_{0}k} \right) \qquad (n\in \mathbb{Z}^{+}) \ .
\end{eqnarray}
Evaluating \ $d^{2}L_{k}(q)/dq^{2}$ \ at \ $q = q_{n}$ \ given in (\ref{B5}) it follows that $q_{2n}$ ($q_{2n+1}$) are maximizers (minimizers) of $L_{k}(q)$. Moreover, using the approximate values of $q_{n}$ given in (\ref{B5}) and approximating sin$\left[ 2/(4\pi \chi_{0}k) \right]$ and cos$\left[ 2/(4\pi \chi_{0}k) \right]$ by their Taylor polynomials of degree $4$ centred at $0$, it follows from (\ref{B1}) that
\begin{eqnarray}
\label{B8}
L_{k}(q_{2n}) &\simeq& \sqrt{\frac{2}{\pi}}(4\pi\chi_{0}k) \ , \cr
&& \cr
L_{k}(q_{2n+1}) &\simeq& \sqrt{\frac{2}{\pi}}\left( \frac{1}{4\pi\chi_{0}k} \right) \times \cr
&& \times\left[ 1+  \frac{2}{\left(4\pi\chi_{0}k\right)^{2}}  -  \frac{1}{\left( 4\pi\chi_{0}k\right)^{4}}  \right]^{-1/2}  \ , \cr
&\simeq& \sqrt{\frac{2}{\pi}}\left( \frac{1}{4\pi\chi_{0}k} \right) \ , \ \
\end{eqnarray}
for each \ $n\in \mathbb{Z}^{+}$. In the approximate equality of $L_{k}(q_{2n})$ and the last approximate equality of $L_{k}(q_{2n+1})$ terms of order $1/(4\pi \chi_{0}k)^{2}$ and smaller have been neglected with respect to $1$. Also, using (\ref{B5}) it follows that the distance between consecutive maximizers is \ $q_{2(n+1)}-q_{2n} \simeq \pi/k$ \ with $n\in\mathbb{Z}^{+}$, which corresponds to half the wavelength \ $\lambda/2 = \pi/k$. Notice that the maximum (minimum) of $L_{k}(q)$ increases (decreases) as the transparency of the mirror decreases.

To end this appendix we show that $L_{k}(q)^{2}$ can be approximated by a Lorentzian in a neighborhood of each maximizer $q_{2n}$ and we comment on the relation of $q_{2n}$ with the cavity resonance frequencies. 

In the following we take
\begin{eqnarray}
\label{B21}
\xi &\equiv& 4\pi \chi_{0}k \ , \cr
&& \cr
v &\equiv& k(q-q_{2n}) \qquad \mbox{for some fixed} \ n\in\mathbb{Z}^{+} \ , \cr
&& \cr
f_{0}(q) &\equiv& 1 +\frac{\xi^{2}}{2} - \xi\mbox{sin}(2kq) - \frac{\xi^{2}}{2}\mbox{cos}(2kq) \ .
\end{eqnarray}
Approximating $f_{0}(q)$ by its second order Taylor polynomial centred at $q_{2n}$, approximating $q_{2n}$ by the value on the right-hand side of (\ref{B5}), approximating sin$(2/\xi)$ and cos$(2/\xi)$ by their fourth order Taylor polynomials centred at $0$, simplifying, and then neglecting terms of order $1/\xi^{2}$ and smaller  with respect to $1$ it follows that  
\begin{eqnarray}
\label{B21bis}
f_{0}(q) &\simeq& \frac{1}{\xi^{2}} + \xi^{2}v^{2} \ .
\end{eqnarray}
Substituting (\ref{B21bis}) in the formula for $L_{k}(q)$ one obtains the following Lorentzian approximation to $L_{k}(q)^{2}$:
\begin{eqnarray}
\label{B22}
L_{k}(q)^{2} &=& \frac{2}{\pi}f_{0}(q)^{-1} \ \simeq \ \frac{2}{\pi\xi^{2}}\left( \frac{1}{v^{2} + \frac{1}{\xi^{4}}} \right) \ .
\end{eqnarray}
Recall that we have been working under the assumption given in (\ref{i1bis}). Therefore, the approximation in (\ref{B5}) is accurate, the fourth order Taylor polynomials of sin$(2/\xi)$ and cos$(2/\xi)$ are accurate approximations of the corresponding functions, and neglecting terms of order \ $1/\xi^{2}$ \ with respect to $1$ is also an accurate approximation. It then follows that the Lorentzian approximation to $L_{k}(q)^{2}$ will be accurate when the error in approximating $f_{0}(q)$ by its second order Taylor polynomial centred at $q_{2n}$ is small, that is, when
\begin{eqnarray}
\label{B26}
\left\vert \frac{1}{2}f_{0}''(q_{2n})\left( \frac{v}{k} \right)^{2} \right\vert &\gg&
\left\vert \frac{1}{3!}f_{0}'''(q^{(0)})\left( \frac{v}{k} \right)^{3} \right\vert  \ ,
\end{eqnarray}
with $q^{(0)}$ in the open interval whose end points are $q$ and $q_{2n}$. 

Simplifying (\ref{B26}) and bounding $\vert f_{0}'''(q^{(0)}) \vert$ one concludes that the Lorentzian approximation to $L_{k}(q)^{2}$ is accurate if (\ref{i1bis}) holds and
\begin{equation}
\label{B27}
\vert v \vert \ \ll \ \frac{3}{2} \ . 
\end{equation} 
Using the definition of $v$ in (\ref{B21}) one obtains that (\ref{B27}) can be written as \ $\vert q - q_{2n} \vert \ll 3/(2k)$. Since \ $\vert q_{2n\pm 1} -q_{2n} \vert = \pi/(2k)$ \ (see (\ref{B5})), it follows that (\ref{i1bis}) and (\ref{B27}) simply state that the Lorentzian approximation to $L_{k}(q)^{2}$ is accurate if the transparency of the movable mirror is small and one restricts to an interval around $q_{2n}$ whose endpoints are not near the minimizers $q_{2n\pm 1}$ of $L_{k}(q)$. 

In the rest of this appendix assume that the condition in (\ref{B27}) is also satisfied. Then the Lorentzian approximation of $L_{k}(q)^{2}$ in (\ref{B22}) is accurate and it has a half-width-at-half-maximum (HWHM) of \ $1/\xi^{2}$. Notice that the value of the HWHM is independent of the value of $n$ in $q_{2n}$ and it decreases as the transparency of the mirror decreases. Moreover, observe that \ $v \in [-1/\xi^{2},1/\xi^{2}]$ \ if and only if \ $kq \in[kq_{2n}-1/\xi^{2},kq_{2n}+1/\xi^{2}]$.

First assume that \ $kq \in [kq_{2n}-1/\xi^{2},kq_{2n}+1/\xi^{2}]$. From (\ref{B8}) and (\ref{B22}) it follows that $L_{k}(q)$ is very large and the mode $V_{k}(x,q)$ in (\ref{7Modos}) is much larger inside the cavity (that is, \ $0 \leq x \leq q = q_{2n}$) than outside of it (that is, \ $q = q_{2n} < x$). Therefore, \ $\omega = ck$ \ coincides with one of the cavity resonance frequencies for these values of $q$. 

Now assume that \ $kq \not\in [kq_{2m}-1/\xi^{2},kq_{2m}+1/\xi^{2}]$ \ for all \ $m\in\mathbb{Z}^{+}$. Then $L_{k}(q)$ is very small and the mode $V_{k}(x,q)$ in (\ref{7Modos}) is much smaller inside the cavity than outside of it. In this case \ $\omega = ck$ \ is very different from any of the cavity resonance frequencies.


\section*{APPENDIX II}

In this appendix we derive various properties of the force $f_{\mbox{\tiny RWA}}(x)$ and the potential $V_{\mbox{\tiny RWA}}(x)$ given respectively in (\ref{81}) and (\ref{PotencialRWA}).

First consider the function
\begin{eqnarray}
\label{i}
f(x) &=& 1 + \xi^{2}\mbox{sin}^{2}(x) - \xi \mbox{sin}(2x) \qquad (x \geq 0).
\end{eqnarray}
Notice that \ $f(x)>0$ \ for all \ $x\in\mathbb{R}$. In the following assume that
\begin{eqnarray}
\label{i1}
\xi \ \gg \ 1 \ .
\end{eqnarray}
From (\ref{NuevoTransparencia2}) it follows that the condition in (\ref{i1}) implies that the transparency of the movable mirror is very small. 

From (\ref{i}) it follows that
\begin{eqnarray}
\label{i2}
f'(x_{n}) = 0 &\Leftrightarrow& \mbox{tan}(2x_{n}) = \frac{2}{\xi} \ .
\end{eqnarray} 
A first approximation to the solutions $x_{n}$ of the equation on the right of (\ref{i2}) can be obtained by using (\ref{i1}) to neglect the term $2/\xi$. One obtains \ $x_{n} \simeq n\pi/2$ \ with \ $n\in\mathbb{Z}^{+}$ \ and $\mathbb{Z}^{+}$ the set of non-negative integers. Using Newton's method \cite{Burden} one can obtain a second more accurate approximation. After one iteration the result is
\begin{eqnarray}
\label{i3}
x_{n} \ \simeq \ n\frac{\pi}{2} + \frac{1}{\xi} \qquad (n\in \mathbb{Z}^{+}) \ .
\end{eqnarray} 
Evaluating numerically the solutions of the equation on the left of (\ref{i2}) one finds that the values on the right in (\ref{i3}) are a very good approximation to the corresponding exact critical point $x_{n}$ if \ $\xi \geq 5$, since the relative error \ $|x_{n}-(n\pi/2+1/\xi)|/|x_{n}|$ \ is less than $10^{-2}$ if \ $\xi \geq 5$ \ and less than $10^{-3}$ if \ $\xi \geq 10$. Also notice that $x_{n}$ coincides with $kq_{n}$ in (\ref{B5}). 

Evaluating $f''(x)$ at $x=x_{n}$ and using the approximation in (\ref{i3}) one obtains that $x_{2n+1}$ ($x_{2n}$) are maximizers (minimizers) of $f(x)$. Since
\begin{eqnarray}
\label{i6}
f_{\mbox{\tiny RWA}}(x) &=& -\frac{1}{2}\left[ 1- \frac{1}{f(x)} \right] \ ,
\end{eqnarray}
it follows that $x_{2n}$ ($x_{2n+1}$) is a maximizer (minimizer) of $f_{\mbox{\tiny RWA}}(x)$ for each \ $n\in\mathbb{Z}^{+}$. Substituting (\ref{i3}) in $f_{\mbox{\tiny RWA}}(x)$ and preserving only terms of order $\xi^{2}$ and $1$ (terms of order $\xi^{-2}$ and smaller are neglected) it follows that
\begin{eqnarray}
\label{i7}
f_{\mbox{\tiny RWA}}(x_{n}) &\simeq&
\begin{cases}
\frac{\xi^{2}}{2}-\frac{7}{18} & \mbox{if $n$ is even} \ , \cr
-\frac{1}{2} & \mbox{if $n$ is odd} \ . \cr
\end{cases}
\end{eqnarray}

In a similar way, one can also calculate the zeros of $f_{\mbox{\tiny RWA}}(x)$ using (\ref{i1}) and Newton's method. One obtains that \ $f_{\mbox{\tiny RWA}}(x) = 0$ \ if and only if \ $x= x_{n}^{*}$ \ or \ $x= x_{n}^{**}$ \ with
\begin{eqnarray}
\label{i8}
x_{n}^{*} = n\pi \qquad \mbox{and} \qquad x_{n}^{**} \simeq n\pi + \frac{2}{\xi}  \ \ \ (n \in \mathbb{Z}^{+}). \ \
\end{eqnarray}
We remark that $n\pi$ is an exact zero of $f_{\mbox{\tiny RWA}}(x)$.

From (\ref{i3}), (\ref{i7}), and (\ref{i8}) it follows that $f_{\mbox{\tiny RWA}}(x)$ has maximizers located at $x_{2n}$ $(n\in\mathbb{Z}^{+})$ and that 
\begin{eqnarray}
\label{i8bis}
x_{n}^{*} = n\pi \ \leq \ x_{2n} \simeq n\pi + \frac{1}{\xi} \ \leq \ x_{n}^{**} \simeq n\pi + \frac{2}{\xi} \ . 
\end{eqnarray}
Notice that the length $2/\xi$ of the interval in (\ref{i8bis}) becomes smaller as the transparency of the movable mirror decreases.

We now show how $f_{\mbox{\tiny RWA}}(x)$ can be approximated by a Lorentzian in a neighborhood of each of its maximizers $x_{2n}$. In the following take \ $n\in\mathbb{Z}^{+}$ \ and
\begin{eqnarray}
\label{hoP1}
u \ = \ x - \left( n\pi + \frac{1}{\xi} \right) \ .
\end{eqnarray}

Using the Taylor series expansion of \ $\mbox{tan}(x)$ \ centred at $0$ it follows that
\begin{eqnarray}
\label{hoP2}
\mbox{tan}(x) \ \simeq \ u + \frac{1}{\xi} \ .
\end{eqnarray}
The right-hand side of (\ref{hoP2}) will be an accurate approximation of the left-hand side if the first correction to the right-hand side is much smaller than \ $(u + \xi^{-1})$, that is, if \ $1\gg 3^{-1}(u+\xi^{-1})^{2}$. Using (\ref{hoP2}) in the expression (\ref{PotencialRWA}) for $V_{\mbox{\tiny RWA}}(x)$ and neglecting $1$ with respect to $\xi^{2}$ in the factor $(1+\xi^{2})$ (recall that we have been working all this time under the assumption in (\ref{i1})) it follows that
\begin{eqnarray}
\label{hoP4} 
V_{\mbox{\tiny RWA}}(x) &\simeq& \frac{1}{2}\left[ u +\left( n\pi + \frac{1}{\xi} \right) \right] -\frac{1}{2}\left[ \mbox{tan}^{-1}(\xi) + n\pi \right] \cr
&& \qquad  -\frac{1}{2}\mbox{tan}^{-1}(\xi^{2}u) \ ,
\end{eqnarray}
for \ $(2n-1)\pi/2 \leq x \leq (2n+1)\pi/2$ \ and \ $x\geq0$. Taking the derivative with respect to $x$ one obtains that
\begin{eqnarray}
\label{hoP5}
f_{\mbox{\tiny RWA}}(x) \ \simeq \ -\frac{1}{2} + \frac{1}{2\xi^{2}}\frac{1}{u^{2} + \xi^{-4}} \ ,
\end{eqnarray}
for \ $(2n-1)\pi/2 < x < (2n+1)\pi/2$ \ and \ $x > 0$ \ and $u$ given in (\ref{hoP1}). One should expect the approximations in (\ref{hoP4}) and in (\ref{hoP5}) to be accurate if  \ $1\gg 3^{-1}(u+\xi^{-1})^{2}$ \ and \ $\xi \gg 1$. Calculating numerically the relative error between $f_{\mbox{\tiny RWA}}(x)$ and the approximation in (\ref{hoP5}) we found that the displaced Lorentzian approximates $f_{\mbox{\tiny RWA}}(x)$ well if \ $\xi \gtrsim 10$ \ and \ $-1/\xi < u < 1/\xi$ \ (that is, $x$ is not very near the points where $f_{\mbox{\tiny RWA}}(x)$ is zero). From (\ref{hoP5}) it follows that the maximizer \ $x_{2n} \simeq (n\pi +\xi^{-1})$ \ of $f_{\mbox{\tiny RWA}}(x)$ has a half-width-at-half-maximum (approximately) equal to $1/\xi^{2}$. Indeed, it can be shown that
\begin{eqnarray}
\label{hoRWA9}
f_{\mbox{\tiny RWA}}\left( n\pi + \frac{1}{\xi} \pm \frac{1}{\xi^{2}} \right) 
&\simeq& \frac{\xi^{2}}{4}  \ \simeq \ \frac{1}{2}f_{\mbox{\tiny RWA}}( x_{2n} ) \ . \ \ \ \ 
\end{eqnarray}
The first approximation in (\ref{hoRWA9}) was obtained by preserving only the terms of order $\xi^{2}$ (terms of order $\xi$ and smaller were neglected). For the second approximation we used (\ref{i7}). 

From (\ref{ii4}) one has \ $f_{\mbox{\tiny RWA}}(x) = -(dV_{\mbox{\tiny RWA}}/dx)(x)$. Hence, the critical points $x_{n}^{*}$ and $x_{n}^{**}$ of $V_{\mbox{\tiny RWA}}(x)$ are given in (\ref{i8}). Evaluating \ $(d^{2}V_{\mbox{\tiny RWA}}/dx^{2})(x)$ \ at the critical points it follows that \ $x_{n}^{*} = n\pi$ \ ($x_{n}^{**} \simeq n\pi + 2/\xi$) are maximizers (minimizers) of $V_{\mbox{\tiny RWA}}(x)$. Moreover, for \ $n\in\mathbb{Z}^{+}$ \ one has
\begin{eqnarray}
\label{iii3}
V_{\mbox{\tiny RWA}}(n\pi) &=& 0 \ , \cr
&& \cr
V_{\mbox{\tiny RWA}}\left( n\pi + \frac{2}{\xi} \right) &\simeq& -\mbox{tan}^{-1}(\xi) \ . 
\end{eqnarray}
The first equality in (\ref{iii3}) is exact, while the second is obtained by neglecting terms of order $1/\xi$ and smaller. Also, one has \ $V_{\mbox{\tiny RWA}}(n\pi + 2/\xi) \rightarrow -\pi/2$ \ if \ $\xi \rightarrow + \infty$. Notice that \ $V_{\mbox{\tiny RWA}}(n\pi) = 0$ \ so that $V_{\mbox{\tiny RWA}}(x)$ has a jump discontinuity at $n\pi$ in the limit $\xi \rightarrow +\infty$.

To end this appendix we drop the assumption \ $\xi \gg 1$ \ in (\ref{i1}) and we prove that \ $-\pi/2 < V_{\mbox{\tiny RWA}}(x) \leq 0$ \ for all \ $x \geq 0$. 

First observe that the inequality \ $V_{\mbox{\tiny RWA}}(x) \leq 0$ \ for all \ $x\geq 0$ \ is obtained by noticing that \ $V_{\mbox{\tiny RWA}}(x)$ \ is a continuous periodic function with maximizers $x_{n}^{*}$ given in (\ref{i8}) and with maximum values given in (\ref{iii3}).

We now consider the lower bound \ $-\pi/2 < V_{\mbox{\tiny RWA}}(x)$ \ for all \ $x \geq 0$. From the definition of $V_{\mbox{\tiny RWA}} (x)$ in (\ref{PotencialRWA}) and from (\ref{iii3}) one has
\begin{eqnarray}
\label{Cota1bis}
V_{\mbox{\tiny RWA}}\left[ (2m+1)\frac{\pi}{2} \right]  &=& - \frac{1}{2}\mbox{tan}^{-1}(\xi) \ \geq \ -\frac{\pi}{4} \ , \cr
&& \cr
V_{\mbox{\tiny RWA}}( m\pi ) &=& 0 \ ,
\end{eqnarray}
for each \ $m \in \mathbb{Z}^{+}$. Therefore, all that remains is to show that \ $-\pi/2 < V_{\mbox{\tiny RWA}}(x)$ \ for all \ $x \geq 0$ \ with \ $x\not= m\pi, \ (2m+1)\pi/2$ \ and \ $m \in \mathbb{Z}^{+}$.

To see this fix \ $x\geq 0$ \ and \ $m\in\mathbb{Z}^{+}$ \ with \ $(2m-1)\pi/2 < x < (2m+1)\pi/2$ \ and \ $x\not= m\pi$. Then one can consider $V_{\mbox{\tiny RWA}}(x)$ to be a function of $\xi$ and we write $V_{\mbox{\tiny RWA}}(x,\xi)$ for clarity. From the definition of $V_{\mbox{\tiny RWA}}(x)$ in (\ref{PotencialRWA}) one has for \ $\xi \geq 0$ \ that
\begin{eqnarray}
\label{Cota1}
\frac{\partial}{\partial \xi} V_{\mbox{\tiny RWA}}(x,\xi) &=& -\frac{1}{2}\frac{1}{\xi^{2}+1} \cr
&& - \frac{1}{2}\frac{2\xi \mbox{tan}(x)-1}{\left[ (1+ \xi^{2})\mbox{tan}(x) -\xi \right]^{2} + 1} \ . \ \ \ \ \
\end{eqnarray}
It then follows that
\begin{eqnarray}
\label{Cota2}
\frac{\partial}{\partial \xi} V_{\mbox{\tiny RWA}}(x,\xi) \ < \ 0 \qquad \mbox{for} \ \ \ \xi \geq 0 \  . \ \ \
\end{eqnarray}
Hence, $V_{\mbox{\tiny RWA}}(x,\xi)$ is a strictly decreasing function of \ $\xi \geq 0$ \ and one has
\begin{eqnarray}
\label{Cota3}
\lim_{\xi \rightarrow + \infty} V_{\mbox{\tiny RWA}}(x,\xi) \ < \ V_{\mbox{\tiny RWA}}(x,\xi)  \  . \ \ \ 
\end{eqnarray}
Also, using $\Theta(y)$ to denote the Heaviside step function equal to $1$ ($0$) if \ $y>0$ \ ($y<0$), it can be shown that
\begin{eqnarray}
\label{Cota4}
\lim_{\xi \rightarrow + \infty} V_{\mbox{\tiny RWA}}(x,\xi) &=& \frac{x}{2} - m\frac{\pi}{2} -\frac{\pi}{2}\Theta \left[ \mbox{tan}(x)\right] \ > \ - \frac{\pi}{2} \ ,\cr
&&
\end{eqnarray}
From (\ref{Cota3}) and (\ref{Cota4}) one obtains that \ $-\pi/2 < V_{\mbox{\tiny RWA}}(x,\xi)$. Therefore, \ $-\pi/2 < V_{\mbox{\tiny RWA}}(x)$ \ for all \ $x \geq 0$.


\section*{APPENDIX III}

In this appendix we take the parameters from \cite{Mar2} and adapt them to our system. It is important to note that the experimental set-up of \cite{Mar2} is dominated by bolometric forces (that is, light absorption deflecting the movable mirror), while the system we are studying only takes into consideration radiation pressure. The intention is to use the parameters of \cite{Mar2} so that one can get an idea of the order of magnitude of the quantities involved in the model.

The movable mirror of \cite{Mar2} is a gold-coated atomic force microscopy cantilever of length \ $L = 223 \mu$m, width \ $W = 22 \mu$m, thickness \ $\delta_{\mbox{\tiny thick}} = 512$ nm, and a spring constant \ $K_{\mbox{\tiny ho}} = 0.01$ N/m. Here we have added to $\delta_{\mbox{\tiny thick}}$ the thickness of the gold layer. For a wavelength of \ $\lambda = 633$ nm \ the movable mirror has a reflectivity of $0.91$. Moreover, the cantilever's fundamental mechanical mode has a frequency of \ $\omega_{1} = 2\pi \times 8.7$ kHz \ and a damping rate of \ $\Gamma_{1} = 30$ Hz.  

From the parameters above it follows that the effective mass $M$ of the gold-coated cantilever is
\begin{eqnarray}
\label{AIII1}
M &=& \frac{K_{\mbox{\tiny ho}}}{\omega_{1}^{2}} \ = \ 3.3 \times 10^{-12} \ \mbox{kg} \ .
\end{eqnarray}
Even if the set-up of \cite{Mar2} were dominated by radiation pressure, the model of this article would not be applicable since \ $\delta_{\mbox{\tiny thick}} \sim \lambda$. To apply the model one would need \ $\delta_{\mbox{\tiny thick}} \ll \lambda$, say \ $\delta_{\mbox{\tiny thick}} = 10^{-2}\lambda = 6.33$ nm. Therefore, we take the mass per unit area $M_{0}$ to be the value that would be obtained if the gold-coated cantilever had uniform mass density and thickness $100$ times smaller, that is,  
\begin{eqnarray}
\label{AIII2}
M_{0} \ = \ \frac{1}{100}\frac{M}{L W} \ = \ 6.8 \times 10^{-6} \ \frac{\mbox{kg}}{\mbox{m}^{2}} \ ,
\end{eqnarray}

From the parameters above we take
\begin{eqnarray}
\label{AIII3}
k_{\scriptscriptstyle{N}}^{\scriptscriptstyle{0}} &=& \frac{2\pi}{\lambda} \ = \ \frac{2\pi}{633} \times 10^{9} \ \mbox{m}^{-1} \ , \cr
\omega_{0} &=& ck_{\scriptscriptstyle{N}}^{\scriptscriptstyle{0}} \ \simeq \ 3 \times 10^{15} \ \mbox{s}^{-1} \ ,
\end{eqnarray}
for the wave-number and angular frequency of the monochromatic laser. Using (\ref{AIII2}) and (\ref{AIII3}) it follows from (\ref{76}) that 
\begin{eqnarray}
\label{AIII4}
\Delta &=& 3.8 \times 10^{12} \times g_{0} \ , \qquad \Omega \ = \ \frac{1562.6}{g_{0}} \ .
\end{eqnarray}
Here $g_{0}$ and $\Delta$ have units of N$^{1/2}$ and 1/s, respectively, while $\Omega$ is non-dimensional. Using (\ref{AIII4}) one then obtains the non-dimensional quantities
\begin{eqnarray}
\label{AIII5}
\Gamma &=& \frac{\Gamma_{1}}{\Delta} \ = \ \frac{7.9 \times 10^{-12}}{g_{0}} \ , \cr
\omega_{\mbox{\tiny ho}} &=& \frac{\omega_{1}}{\Delta} \ = \ \frac{1.4 \times 10^{-8}}{g_{0}} \ .
\end{eqnarray}
From (\ref{AIII4}) and (\ref{AIII5}) it follows that
\begin{eqnarray}
\label{NuevoAIII}
\frac{\Gamma}{\Omega} \sim 10^{-14} , \qquad \frac{\omega_{\mbox{\tiny ho}}}{\Omega} \sim 10^{-11} , \qquad \frac{\Gamma}{\omega_{\mbox{\tiny ho}}} \sim 10^{-3} . 
\end{eqnarray}

According to \cite{Nussenzveig} the transmissivity $T$ of the \textit{delta}-mirror is related to \ $\xi = 4\pi \chi_{0}k_{\scriptscriptstyle{N}}^{\scriptscriptstyle{0}}$ \ in (\ref{76}) by 
\begin{eqnarray}
\label{AIII6}
T \ = \ \frac{1}{1+ \left( \frac{\xi}{2} \right)^{2}} \ \ &\Leftrightarrow& \ \ \xi \ = \ 2\sqrt{\frac{R}{1-R}} \ .
\end{eqnarray}
Here we used that \ $R + T = 1$ \ with $R$ the reflectivity. Using the value \ $R = 0.91$ \ given above one obtains that \ $\xi = 6.4$. Also, \ $\xi = 50$ \ if and only if \ $R = 0.9984$.

All that remains is to give a typical value for $g_{0}$. We determine this by calculating the average laser power incident on the movable mirror.

Using the parameters above we take the cavity to be given by the box
\begin{eqnarray}
\label{AIII7}
V &=& \left[ 0,q(t) \right] \times \left[ -\frac{L}{2}, \frac{L}{2} \right] \times \left[ -\frac{W}{2}, \frac{W}{2} \right] \ .
\end{eqnarray}
Moreover, the movable mirror is given by the plane \ $\mathcal{P} = \{ q(t) \} \times \left[ -\frac{L}{2}, \frac{L}{2} \right] \times \left[ -\frac{W}{2}, \frac{W}{2} \right]$.

Using (\ref{Vkt}) and (\ref{FormaCompleja}) one can identify the part of the complex-valued vector potential \ $2g_{0}\tilde{V}_{k_{\scriptscriptstyle{N}}^{\scriptscriptstyle{0}}}[x,q(t)]e^{-i\omega_{0}t}$ \ associated with the laser:
\begin{eqnarray}
\label{Alaser}
A_{0L}(x,t) &=& i\frac{2g_{0}}{\sqrt{2\pi}}e^{-i(k_{\scriptscriptstyle{N}}^{\scriptscriptstyle{0}}x + \omega_{0}t)} \ .
\end{eqnarray} 
Using (\ref{Campos}) it follows that the electromagnetic field associated with the laser is given by
\begin{eqnarray}
\label{CamposLaser}
\mathbf{E}_{L}(x,t) &=& -\mathbf{z}\left(\frac{2}{\pi}\right)^{1/2}g_{0}k_{\scriptscriptstyle{N}}^{\scriptscriptstyle{0}}e^{-i(k_{\scriptscriptstyle{N}}^{\scriptscriptstyle{0}}x+\omega_{0}t)} \ , \cr
\mathbf{B}_{L}(x,t) &=& -\mathbf{y}\left(\frac{2}{\pi}\right)^{1/2}g_{0}k_{\scriptscriptstyle{N}}^{\scriptscriptstyle{0}}e^{-i(k_{\scriptscriptstyle{N}}^{\scriptscriptstyle{0}}x+\omega_{0}t)} \ .
\end{eqnarray}
Hence, the time-averaged Poynting vector associated with the laser is
\begin{eqnarray}
\label{SLaser}
\mathbf{S}_{L}(x,t) &=& \frac{c}{8\pi}\mathbf{E}_{L}(x,t)\times \mathbf{B}_{L}(x,t)^{*} \ , \cr
&=& -\frac{1}{c}\left(\frac{g_{0}\omega_{0}}{2\pi}\right)^{2} \mathbf{x} \ . 
\end{eqnarray}
It follows that the average laser power incident on the movable mirror from the right is given by
\begin{eqnarray}
\label{AIII8}
P  \ = \ \int_{\mathcal{P}} \mathbf{S}_{L}(x,t)\cdot (-\mathbf{x}) da  
\ = \ \frac{1}{c}\left( \frac{g_{0}\omega_{0}}{2\pi} \right)^{2}LW \ . \ \
\end{eqnarray}
Let $P_{\mbox{\tiny max}}$ be the maximum average incident laser power. It follows from (\ref{AIII8}) that 
\begin{eqnarray}
\label{AIII10}
g_{0} &\leq& \frac{2\pi}{\omega_{0}}\sqrt{\frac{cP_{\mbox{\tiny max}}}{LW}} \ . \qquad
\end{eqnarray}
If the maximum laser power is $P_{\mbox{\tiny max}} = 1$ Watt, it follows from (\ref{AIII10}) and the values of $L$, $W$, and $\omega_{0}$ above that \ $g_{0} \leq 5.2 \times 10^{-7} \ \mbox{N}^{1/2}$ where $g_{0}$ is given in units of N$^{1/2}$. Taking \ $g_{0} \leq 10^{-6}$ N$^{1/2}$ one gets from (\ref{AIII4}) the typical values \ $\Delta \leq 3.8 \times 10^{6}$ 1/s \ and \ $\Omega \geq 10^{9}$.


\end{document}